\documentclass[oribibl]{llncs}

\usepackage{amssymb}
\usepackage{graphicx}
\usepackage{comment}
\usepackage{amsmath}
\usepackage{color}
\usepackage{url}
\usepackage{cite}
\usepackage{caption}
\usepackage{algorithm}
\usepackage{algpseudocode}
\usepackage{wrapfig}

\makeatletter
\newcommand{\xRightarrow}[2][]{\ext@arrow 0359\Rightarrowfill@{#1}{#2}}
\makeatother

\urldef{\mailsa}\path|{mmilani, bertossi}@scs.carleton.ca|

\newcommand{\vectt}[1]{\bar{#1}}

\newcommand{\mc}[1]{\mathcal{ #1}}

\newcommand{\nit}[1]{{\it #1}}
\newcommand{\boxtheorem}{\hfill $\blacksquare$\vspace*{0.2cm}}
\newcommand{\ignore}[1]{}
\newcommand{\defproof}[2]{{\noindent\bf Proof of #1:\
}#2 \boxtheorem\\ \vspace{2mm}}

\newcommand{\da}{Datalog~}
\newcommand{\dpm}{{Datalog}$^\pm$}
\newcommand{\de}{{Datalog}$^\exists$}
\newcommand{\dplus}{{Datalog}$^+$}

\newcommand{\m}{\;\!\!}

\newcommand{\ms}{{\sf MagicD}$^+$}
\newcommand{\tgd}{TGD}
\newcommand{\SCh}{{\em S\m{}Ch}}
\newcommand{\WSCh}{{\em WS\m{}Ch}}
\newcommand{\WS}{{\em WS}}
\newcommand{\WA}{{\em WA}}
\newcommand{\JWS}{{\em JWS}}
\newcommand{\JA}{{\em JA}}
\newcommand{\GSCh}{{\em GS\m{}Ch}}
\newcommand{\QAA}{{\sf SChQA}}

\newcommand{\edg}{{\em EDG}}
\newcommand{\dg}{{\em DG}}

\newcommand{\SIPS}{{\em SIPS}}

\newcommand{\fp}[1]{\nit{Fin\!P\!\m{}o\m{}ss}(#1)}

\begin{document}
\thispagestyle{empty}
\pagestyle{plain}

\mainmatter  

\title{Extending Weakly-Sticky \dpm: Query-Answering Tractability and Optimizations}

\titlerunning{Extending Weakly-Sticky \dpm: Query-Answering Tractability and Optimizations}

\author{{\bf Mostafa Milani} \and {\bf Leopoldo Bertossi}}
%
\authorrunning{M. Milani \and L. Bertossi}

\institute{Carleton University, School of Computer Science\\
Ottawa, Canada}

%
%

\maketitle

\vspace{-6mm}
\begin{abstract}
{\em Weakly-sticky} (\WS\;) \dpm \ is an expressive member of the family of \dpm \ programs that is based on the syntactic notions of {\em stickiness} and {\em weak-acyclicity}. Query answering over the \WS \ programs has been investigated, but there is still much work to do on  the design and implementation of practical query answering (QA) algorithms and their optimizations. Here, we study sticky and \WS \ programs from the point of view of the behavior of the chase procedure,  extending the stickiness property of the chase to that of {\em generalized stickiness of the chase (gsch-property)}. With this property we specify the semantic class of \GSCh \ programs, which includes sticky and \WS \ programs, and other syntactic subclasses that we identify. In particular, we introduce  {\em joint-weakly-sticky} (\JWS\;) programs, that include \WS \ programs. We also propose a bottom-up QA algorithm for a range of subclasses of \GSCh. The algorithm runs in polynomial time (in data) for \JWS \  programs.  Unlike the \WS \ class, \JWS \ is closed under a general magic-sets rewriting procedure  for the optimization of programs with existential rules. We apply the magic-sets rewriting in combination with the proposed QA algorithm
 for the optimization of QA over \JWS \ programs.
\end{abstract}

\vspace{-9mm}
\section{Introduction}

\vspace{-2mm}
Ontology-based data access (OBDA) \cite{poggi} allows to access, through a conceptual layer that takes the form of an ontology, underlying data that is usually stored in a relational database. Queries can be expressed in terms of the ontology language, but are answered by eventually appealing to the extensional data underneath. Common languages of choice for representing ontologies are certain classes (or fragments) of {\em description logic} (DL) \cite{artale} and, more recently, of {\em \dpm} \cite{\ignore{cali09,}cali09-ws,cali12}. Those classes  are expected to be computationally well-behaved in relation to query answering (QA).
 Several approaches for QA, and a number of techniques have been proposed for DL-based  ~\cite{artale,poggi} and  \dpm\!-based OBDA \cite{cali09-ws}.
\ignore{ Since the conceptual layer provided by the ontology can unify possibly heterogeneous data sources, through a shared meaning of the terms of a  vocabulary in common,
OBDA has been applied in several areas, such as virtual data integration \cite{lenzerini02} and the semantic web with RDF data \cite{arenas}.} 
In this work we concentrate on the conjunctive QA problem from relational data through \dpm \ ontologies.

\dpm, as an extension of the \da query language~\cite{ceri}, allows in rule heads (i.e. consequents): existentially quantified variables ($\exists$-variables), equality atoms, and a false propositional atom, say $\mathbf{false}$, to represent ``negative program constraints" \cite{cali09-ws,cali12-is,cali12}. Hence the ``$+$" in \dpm, while the ``$-$" reflects syntactic restrictions on programs for better computational properties.

\dpm \ is expressive enough to represent in logical and declarative terms useful ontologies, in particular those that capture and extend the common
conceptual data models \cite{cali12-is} and Semantic Web data \cite{arenas}.
The rules of a \dpm \ program can be seen as forming an ontology  on top of an extensional database, $D$, which may be {\em incomplete}. In particular, the ontology: (a)   provides a ``query layer" for $D$,
enabling  OBDA, and (b) specifies a completion of $D$.

{\em In the rest of this work we will assume that programs contain only existential rules (plus extensional data). When programs are subject to syntactic restrictions, we talk about \dpm \ programs, whereas when no
conditions are assumed or applied, we talk about \dplus \ programs, also called \de \ programs}~\cite{baget, cali09-ws, leone11, krotzsch}.

From the semantic and computational point of view, the completion of the underlying extensional instance $D$ appeals to so-called {\em chase} procedure that, starting  from $D$, iteratively enforces the rules in the ontology.
That is, when a rule body (the antecedent) becomes true in the instance so far, but not the head (the consequent), a new tuple is generated.  This process may create new values (nulls) or propagate values to the same or other {\em positions}. The latter correspond to the arguments in the schema predicates.

\vspace{-1mm}
\begin{example} \label{ex:intr} Consider a \dpm \ program  $\mc{P}$ with extensional database $D=\{r(a,b)\}$ and  set of rules $\mc{P}^r$:

\vspace{-5mm}

\hspace{-1cm}\begin{minipage}[t]{0.45\textwidth}\centering
\begin{eqnarray}r(X,Y) \! ~\rightarrow~ \! \exists Z \ r(Y,Z).\label{eq:s1}\end{eqnarray}
\end{minipage}\hspace{5mm}
\begin{minipage}[t]{0.45\textwidth}\centering
\begin{eqnarray}r(X,Y),r(Y,Z) \! ~\rightarrow~ \! s(X,Y,Z). \hspace{1cm}\label{eq:s2}\end{eqnarray}
\end{minipage}
\vspace{2mm}



The positions for this schema are: $r[1],r[2],s[1],s[2],s[3]$. The extension of $D$ generated by the chase includes the following tuples (among infinitely many  others): $r(b,\zeta_1),$ $s(a,b,\zeta_1),r(\zeta_1,\zeta_2),s(b,\zeta_2,\zeta_1)$. Notice that $s(a,b,\zeta_1)$ and $s(b,\zeta_1,\zeta_2)$ are obtained by replacing the {\em join variable} $Y$ (i.e. repeated) in the body of (\ref{eq:s2}) by $b$ and $\zeta_1$, resp. \boxtheorem\end{example}\vspace{-3mm}

The result of the chase, seen as an instance for the combined ontological and relational schema, is also called ``the chase".  The chase (instance) extends $D$, but may be infinite;
and
gives the semantics to the \dpm \ ontology, by providing an intended model, and can be used for QA.  At least conceptually, the query can be posed directly to the materialized chase instance. However, this may not be the best way to
go about QA, and computationally better alternatives have to be explored.

 Actually, when the chase is infinite, (conjunctive) QA may be undecidable \cite{johnson}. However, in some cases, even with an  infinite chase, QA is still computable (decidable), and even tractable in the size of $D$. In fact,  syntactically restricted subclasses of \dplus \ programs have been identified and characterized for which
QA is decidable, among them: {\em linear}, {\em guarded} and {\em weakly-guarded}, {\em sticky} and {\em weakly-sticky} (\WS\;)~\cite{cali09-ws,cali12} \dpm\!.

Sticky \dpm \ is a syntactic class of programs characterized by  syntactic restrictions on join variables. \WS \ \dpm \ extends sticky \dpm \ by also capturing the well-known class of {\em weakly-acyclic programs}~\cite{fagin}, which is defined in terms of the syntactic notions of {\em finite-} and {\em infinite-rank} positions. Accordingly, \WS \ \dpm \ is characterized by restrictions on join variables occurring
in infinite-rank positions.
 A non-deterministic QA algorithm  for \WS \ \dpm \ is presented in \cite{cali12}, to establish the theoretical result that QA can be done in polynomial-time in data.

In this work, we concentrate on sticky and \WS \ \dpm\!, because they have found natural applications in our previous work  on extraction of quality data from possible dirty databases~\cite{milani15}. The latter task is accomplished through QA, so that the need for efficient QA algorithms becomes crucial. Accordingly,
the main motivations,  goals, and results (among others) for/in this work are:
\vspace{-1mm}
\begin{itemize}\item[(A)] Providing a practical, bottom-up QA algorithm for \WS \ \dpm\!. Being bottom-up, it is expected
to be based on (a variant of) the chase. Since the latter can be infinite, the query at hand guarantees that the need to generate only an initial, finite portion of the chase.
\item[(B)] Optimizing the QA algorithm through a {\em magic-sets} rewriting technique, to make it more query sensitive.
\end{itemize}
\vspace{-1mm}

For (B), we apply the magic-sets technique for \dplus \ first introduced in~\cite{alviano}, which we denote with \ms. \ Extending classical magic-sets for Datalog \cite{\ignore{bancilhon,beeri-ms,}ceri},
  \ms prevents existential variables from getting bounded, a reasonable adjustment that essentially preserves the semantics of existential rules  during the rewriting.
  Unfortunately, the class of \WS \ \dpm \ programs is provably not closed under \ms, meaning that the result of applying \ms \ to a \WS \ program may not be \WS \ anymore. This led us to search for a more general class of programs that is: (i) closed under \ms, (ii) extends \WS \ \dpm, and (iii) has an efficient QA algorithm. \ Notice that at this point both syntactic and semantic classes may be investigated, and we do so. The latter classes refer to the properties of the chase as
  an instance.

Sticky programs enjoy the {\em stickiness property of the chase}, which -in informal terms- means the following:
 \ If, due to the application of a rule during the chase, a value replaces a join variable in the rule body, then that value is propagated through all the possible subsequent steps, i.e. the value ``sticks". The ``stickiness property of the chase" defines a ``semantic class", \SCh, in the sense that it is characterized in terms of the chase for programs that include an extensional database. This class properly extends sticky \dpm \ \cite{cali12}.

We can relax the condition in the {\em sch-property}, and define the {\em generalized-stickiness property of the chase}. It is as for the {\em sch-property}, but with the propagation condition only on join variables that {\em do not} appear in the {\em finite positions};
the latter being  those where finitely many different values may appear during the chase. With this property we define the new semantic class of {\em GSCh} \ programs.  However, we make notice that, given a program $\mc{P}$ consisting of a set of rules $\mc{P}^r$ and an extensional instance $D$, computing (deciding) $\fp{\mc{P}}$, the set of finite positions of $\mc{P}$, is unsolvable
(undecidable) \cite{deutsch}. Accordingly, it is also undecidable if a Datalog$^+$ program belongs to the {\em GSCh} class.

Starting from the definition of the {\em GSCh} class, we can define, backwardly, a whole range of  different semantic classes between {\em Sticky} and {\em GSCh}, by replacing in the definition of the latter  the
condition on the set of non-finite positions by a stronger one that appeals to a  superset of them. Each of these supersets  is represented through its complement, which is determined  by an abstract {\em selection function} $\mc{S}$ that identifies a set of finite positions. Such a function,  given a program $\mc{P}$, returns a subset $\mc{S}(\mc{P})$ of $\fp{\mc{P}}$ (making $\mc{S}$ sound, but possibly incomplete w.r.t $\fp{\mc{P}}$). $\mc{S}$ may be computable or not, and  may depend  on $\mc{P}^r$ alone or on the combination of $\mc{P}^r$ and $D$. Hence we split $\mc{P}$ into $\mc{P}^r$ and $D$. The corresponding semantic class of programs, those enjoying the {\em $\mc{S}$-stickiness property of the chase}, is denoted with \SCh$(\mc{S})$.


In particular, if $\mc{S}^\top$ is the non-computable function that selects all finite positions, \GSCh \ $=$ \SCh$(\mc{S}^\top)$.  If $\mc{S}^\nit{rank}$ selects the finite-rank positions (that happen to be finite positions) \cite{fagin}, then \WSCh \ $=$ \ \SCh$(\mc{S}^\nit{rank})$ is a new semantic class programs, those  with the {\em weak-stickiness property of the chase}. And for the class \SCh \ of programs we started from above, it holds \SCh \ $=$ \ \SCh$(\mc{S}^\bot)$, with $\mc{S}^{\bot}$ always returning the empty set of positions. Notice that $\mc{S}^\nit{rank}$ and $\mc{S}^\bot$ are both computable, and they do not use the extensional instance $D$, but only the program. In this sense, we say that they are {\em syntactic selection functions}.

\begin{figure}[h]
\centering
\resizebox{!}{3.5cm}{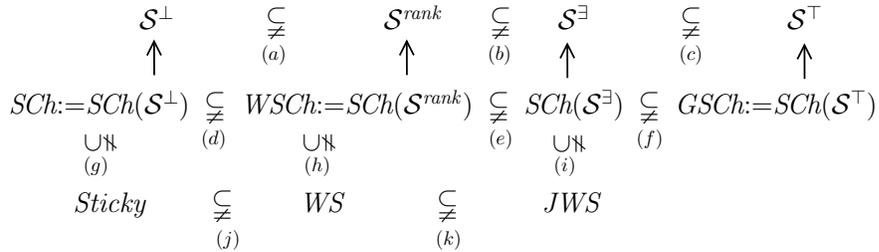}
\vspace{-4mm}
\caption{Semantic and syntactic program classes, and selection functions}
\label{fig:range-classes}
\end{figure}

We can see that the combination of  selection functions with the $\mc{S}$-based notion of stickiness property of the chase (i.e. that only values in join variables in positions outside those selected by $\mc{S}$ propagate all the way through), defines a range of semantic classes of programs starting with \SCh, ending with \GSCh, and with \SCh$(\mc{S}^\nit{rank})$ in between. They are
shown in  ascending order of inclusion, from left to right,   in the middle layer
 of Figure~\ref{fig:range-classes}. There, the upper layer shows the corresponding selection functions ordered by inclusion (of their images).\\


\vspace{-9mm}
\vspace{-8mm}

A parallel and corresponding range of syntactic classes, also ordered
 by set inclusion, is shown in the lower layer. It includes the sticky and \WS \ classes (cf. Figure~\ref{fig:range-classes}, bottom).
 Each syntactic class only partially represents its semantic counterpart, in the sense that the former: does not consider extensional instances, appeals to the same selection function, but also imposes additional
 syntactic conditions on the set of rules.
 All the inclusions in Figure \ref{fig:range-classes} are proper, as examples we provide in this work will show (but (g) and (j) are known \cite{cali12}).

 In this work, our main goal is to introduce and investigate the semantic class \SCh$(\mc{S}^\exists)$, determined by the selection function $\mc{S}^\exists$ that is defined in terms of the {\em existential dependency graph} of a program \cite{krotzsch} (a syntactic, computable construction). We also introduce and investigate its corresponding syntactic class of {\em joint-weakly-sticky (JWS)} programs. The latter happens
 to satisfy desiderata (A) and (B) above. Actually, about (A), we provide for the class \SCh$(\mc{S}^\exists)$  a polynomial-time, chase-based, bottom-up QA algorithm, which can be applied to {\em JWS} (and all its semantic and syntactic subclasses) in particular. This is a general
 situation: The  polynomial-time QA algorithms for the classes  {\em Sticky} \cite{cali12}, {\em WS} \cite{cali12,milani16er}, and {\em JWS} (this work) rely basically
 on the properties of the semantic class rather than on the specific syntactic restrictions. Hence our interest is in investigating the particular semantics classes, and semantic classes in general, as defined
 by selection functions. About (B), notice that if we start with a \WS \ program, we can apply \ms \ to it, obtaining a \JWS \ program, for which QA can be done in polynomial time.

The paper is structured as follows: Section~\ref{sec:preliminaries} is a review of some basics of the database theory, the chase procedure, and \dpm\!. Section~\ref{sec:gsch} contains the definition of the stickiness and general-stickiness properties of the chase and the \SCh \ and \GSCh \ semantic classes. Section~\ref{sec:pclasses} is about the ranges of syntactic and semantic program subclasses of \GSCh. The \JWS \ class of programs is introduced in Section~\ref{sec:jws}. Section~\ref{sec:qa} and Section~\ref{sec:magic} contain the QA algorithm and \ms\!. In this
paper we use mainly intuitive and informal introductions of concepts and techniques, illustrated by examples. The precise technical developments can be found in the Appendices
of \cite{qaMS}. 

\vspace{-3mm}
\section{Preliminaries}
\label{sec:preliminaries}
\vspace{-3mm}


We start with a  relational schema $\mc{R}$ containing two disjoint ``data" sets:  $\mc{C}$, a possibly infinite domain of {\em constants}, and  $\mc{N}$, of  infinitely many {\em labeled nulls}. It also contains predicates of fixed and finite arities. If $p$ is an $n$-ary predicate (i.e. with $n$ arguments) and $1\leq i \leq n$, $p[i]$ denotes its $i$-th position. With $\mc{R}$, $\mc{C}$, $\mc{N}$ we can build a language $\mc{L}$ of first-order (FO) predicate logic, that has $\mc{V}$ as its infinite set of {\em variables}.
We denote with $\vectt{X}$, etc., finite sequences of variables. A {\em term} of the language is a constant, a null, or a variable. An {\em atom} is of the form $p(t_1, \ldots, t_n)$, with $p \in \mc{R}$, $n$-ary predicate, and
$t_1, \ldots, t_n$ terms. An atom is {\em ground}, if it contains no variables. An {\em instance} $I$ for schema $\mc{R}$ is a possibly infinite set of ground atoms. The \emph{active domain} of an instance $I$, denoted
${\it Adom}(I)$, is the set of constants or nulls that appear in $I$.
Instances can be used as interpretation structures for the FO language $\mc{L}$. Accordingly, we can use the notion of formula satisfaction of FO predicate logic.

A conjunctive query (CQ) is a FO formula,  $\mc{Q}(\bar{X})$, of the form: \ $\exists  \bar{Y}(p_1(\vectt{X}_1) \wedge \cdots \wedge p_n(\vectt{X}_n))$, with $\bar{Y} := (\bigcup \vectt{X}_i) \smallsetminus \vectt{X}$.
For an instance $I$, $\vectt{t} \in (\mc{C} \cup \mc{N})^n$ \ is an {\em answer} to $\mc{Q}$ if $I \models \mc{Q}[\bar{t}]$, with $\bar{t}$ replacing the variables in $\bar{X}$. $\mc{Q}(I)$ denotes the set of answers to $\mc{Q}$ in $I$. $\mc{Q}$ is Boolean (a BCQ) when $\vectt{X}$ is empty, and when true in $I$, $\mc{Q}(I) := \{\nit{yes}\}$. Otherwise, $\mc{Q}(I) = \emptyset$. Notice that a CQ can be expressed as a
rule of the form $p_1(\vectt{X}_1),...,p_n(\vectt{X}_n)\rightarrow \nit{ans}_\mc{Q}(\vectt{X})$, where $\nit{ans}_\mc{Q}(\cdot) \notin \mc{R}$ is an auxiliary predicate.  The query answers form the
extension of the answer-collecting predicate $\nit{ans}_\mc{Q}(\cdot)$.\footnote{When $\mc{Q}$  is
Boolean,  $\nit{ans}_{\mc{Q}}$ is a propositional atom; and if $\mc{Q}$ is true in $I$, then $\nit{ans}_\mc{Q}$ can be reinterpreted as the query answer.}

A \emph{tuple-generating dependency} (\emph{\tgd}), also called
\emph{existential rule} or simply a {\em rule} is a sentence, $\sigma$, of $\mc{L}$ of the form: $p_1(\vectt{X}_1), \ldots, p_n(\vectt{X}_n)\rightarrow \exists \vectt{Y} p(\vectt{X},\vectt{Y}),$ \ignore{\label{eq:tgd}}
with $\vectt{X}_i$ indicating the variables appearing in $p_i$ (among possibly elements from $\mc{C}\ignore{ \cup \mc{N}}$), and an implicit universal quantification over all variables
 in $\bar{X}_1, \ldots, \bar{X}_n, \bar{X}$, and $\bar{X} \subseteq \bigcup_i \bar{X}_i$, and the dots in the antecedent standing for conjunctions.\footnote{A query of this form can be
seen and treated as a new TGD containing a fresh head predicate.} The variables in~$\vectt{Y},$ that could be empty, are \emph{existential
  variables}.
 With ${\it
  head}(\sigma)$ and ${\it body}(\sigma)$ we denote the sets of atoms in the consequent and the antecedent of $\sigma$, respectively.  The notions of satisfaction by an instance $I$ of a TGD $\sigma$ (denoted $I \models \sigma$),
  and of a set of TGDs, are defined as in FO logic.

A Datalog$^+$ program $\mc{P}$ consists of a set of rules $\mc{P}^r$ and an extensional database instance $D$, i.e. a finite instance whose atoms contain only elements from $\mc{C}$.  The set of models of $\mc{P}$, denoted by $\nit{Mod}(\mc{P})$, contains all instances $I$, such that $I \supseteq D$ and $I \models \mc{P}^r$. Given a  CQ $\mc{Q}$, the set of answers to $\mc{Q}$ from $\mc{P}$ is defined by $\nit{ans}(\mc{Q}, \mc{P}) := \bigcap_{I
\in \nit{Mod}(\mc{P})} \mc{Q}(I)$.

The \emph{chase} procedure is a fundamental algorithm in different database
problems, including implication of database dependencies, query containment, and
CQ answering under dependencies~\cite{beeri,cali12,fagin,johnson,maier}. For the latter problem
\cite{cali12,fagin}, the idea is that, given a set of
dependencies over a database schema and an instance as input,
the chase enforces the dependencies by adding new tuples into the instance, so that the result satisfies the
constraints (cf. Appendix~B in \cite{qaMS} for more details).

\vspace{-2mm}
\begin{example}\label{ex:chase} (example \ref{ex:intr} cont.) With the given instance $D$ and the assignment  $\theta\!: \ X\mapsto a, Y\mapsto b$, rule (\ref{eq:s1}) is not satisfied: $D \models r(X,Y)[\theta]$, but $D \not \models \exists Z\;r(Y,Z)[\theta]$. Then, the chase inserts a new tuple $r(b,\zeta_1)$ into $D$ ($\zeta_1$ is a fresh null), resulting in instance $D_1$.
$D_1$ does not satisfy (\ref{eq:s2}), so the chase inserts $s(a,b,\zeta_1)$, resulting  in instance $D_2$. The chase continues, without stopping, creating an infinite instance: \ ${\it chase}(\mc{P}) = \{r(a,b),r(b,\zeta_1),s(a,b,\zeta_1),r(b,\zeta_1),r(\zeta_1,\zeta_2),s(b,\zeta_1,\zeta_2), \ldots\}$.
\boxtheorem
\end{example}\vspace{-3mm}

The instance  resulting from  the chase procedure is also  called ``the  chase". As such, it  is a so-called
\emph{universal model}~\cite{fagin}, i.e. a representative of all models in
$\nit{Mod}(\mc{P})$. In particular, the answers to a CQ $\mc{Q}$ under $\mc{P}$, i.e. those in $\nit{ans}(\mc{Q}, \mc{P})$,
can be
computed by evaluating $\mc{Q}$ over the chase (and discarding the answers
containing nulls). 
The chase procedure may not terminate, and it is in general undecidable if it terminates, even for a fixed instance \cite{deutsch}.

Several sufficient conditions, syntactic~\cite{deutsch,fagin,marnette} and data-dependent~\cite{meier}, that guarantee chase termination have been identified. {\em Weak-acyclicity}~\cite{fagin} is one of the former, and is defined using the dependency graph.

\begin{wrapfigure}{r}{0.35\textwidth}
\centering
\vspace{-1.25cm}
\resizebox{!}{2.25cm}{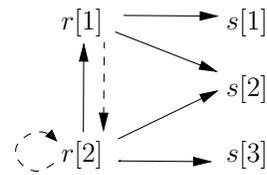}
\vspace{-2mm}
\caption{Dependency graph}\label{fig:dg}
\vspace{-1.0cm}
\end{wrapfigure}

\vspace{-2mm}
\begin{example} \label{ex:dg}  (example \ref{ex:chase} cont.) The {\em dependency graph} (\dg) of $\mc{P}^r$ (cf. Figure~\ref{fig:dg}) is a directed graph whose vertices are the positions of $\mc{R}$. \end{example}

The edges are defined as follows: for every $\sigma \in \mc{P}^r$, $\forall$-variable $X$ in ${\it head}(\sigma)$, and position $\pi$ in ${\it body}(\sigma)$: \ 1. for each occurrence of $X$ in position $\pi'$ in ${\it head}(\sigma)$, create an edge from $\pi$ to $\pi'$. 2. for each $\exists$-variable $Z$ in position $\pi''$ in ${\it head}(\sigma)$, create a {\it special edge} (dashed) from $\pi$ to $\pi''$.

The {\it rank of a position} is the maximum number of special edges over all (finite or infinite) paths ending at that position. $\Pi_F(\mc{P}^r)$ is the set of finite-rank positions in $\mc{P}^r$.  A program is {\em weakly-acyclic} (\WA) if all of the positions have finite-rank. Here, $r[1], r[2] \notin \Pi_F(\mc{P}^r)$, so the program is not \WA.
\boxtheorem

In a program with finite- and infinite-rank positions,
every finite-rank position is finite: For any extensional instance $D$, during the chase only polynomially many  different values appear in them  (in data) \cite{cali12}. However,
in infinite-rank positions, there may be infinitely many values (and the chase does not terminate). In particular, for every \WA \ program and instance $D$ the chase terminates in polynomially many steps with respect to the size of $D$ \cite{fagin}.

The notions of finite and infinite positions mentioned above rely on the chase instance and hence a program's data: Given a program $\mc{P}$ with schema $\mc{R}$, the set of finite positions of $\mc{P}$, that we refer to as $\fp{\mc{P}}$, is the set of positions where finitely many values appear in $\nit{chase}(\mc{P})$. Every position that is not finite is infinite.

Conjunctive query answering w.r.t an arbitrary set of TGDs is in general undecidable \cite{beeri-icalp}. The \dpm \ family is formed by syntactic subclasses of \dplus \ programs that are defined by imposing restrictions on the sets of TGDs rules in a program, to guarantee
decidability, and in several cases, tractability of QA. In this work we concentrate on the sticky and \WS \ classes of programs.

\vspace{-3mm}
\section{Stickiness of the Chase and its Generalization}\label{sec:gsch}
\vspace{-2mm}

The {\em``stickiness property of the chase"} ({\em sch-property}) \cite{cali12} is a ``semantic" property of \dplus \ programs in relation to the way the chase behaves with the extensional data. We informally introduce it here. A program has this property if, due to the application of a rule $\sigma$, when a value replaces a repeated variable in a rule-body, then that value also appears in all the head atoms obtained through the iterative enforcement of applicable rules that starts with $\sigma$'s application. In short, the value is propagated through all possible subsequent chase steps.

\vspace{-2mm}
\begin{example}\label{exp:chs} Consider $\mc{P}_1$ with $D_1 = \{r(a,b)$$,$$r(b,c)\}$, and $\mc{P}_1^r$ containing:\end{example}

\vspace{-9mm}
\[ \arraycolsep=0pt
\begin{array}{rcl c rcl c rcl}
r(X,Y),r(Y,Z) &\;\rightarrow\;& p(Y,Z). &\hspace{5mm}&p(X,Y) &\;\rightarrow\;& \exists Z\;s(X,Y,Z). &\hspace{5mm}& s(X,Y,Z) &\;\rightarrow\;& u(Y).\\
\end{array}
\]

\begin{minipage}[t]{0.5\textwidth}\centering
\begin{minipage}[t]{0.25\textwidth}\centering\vspace{-3.1cm}
\hspace{-3.5cm}\resizebox{!}{2.1cm}{\begin{picture}(0,0)%
\includegraphics{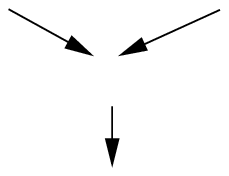}%
\end{picture}%
%
%
\setlength{\unitlength}{3947sp}%
\begingroup\makeatletter\ifx\SetFigFont\undefined%
\gdef\SetFigFont#1#2#3#4#5{%
  \reset@font\fontsize{#1}{#2pt}%
  \fontfamily{#3}\fontseries{#4}\fontshape{#5}%
  \selectfont}%
\fi\endgroup%
\begin{picture}(1480,1275)(458,-726)
\put(473,367){\makebox(0,0)[lb]{\smash{{\SetFigFont{12}{14.4}{\rmdefault}{\mddefault}{\updefault}{\color[rgb]{0,0,0}$r(a,\textbf{b})$}%
}}}}
\put(1773,367){\makebox(0,0)[lb]{\smash{{\SetFigFont{12}{14.4}{\rmdefault}{\mddefault}{\updefault}{\color[rgb]{0,0,0}$r(\textbf{b},c)$}%
}}}}
\put(1182,-106){\makebox(0,0)[lb]{\smash{{\SetFigFont{12}{14.4}{\rmdefault}{\mddefault}{\updefault}{\color[rgb]{0,0,0}$p(\textbf{b},c)$}%
}}}}
\put(1182,-649){\makebox(0,0)[lb]{\smash{{\SetFigFont{12}{14.4}{\rmdefault}{\mddefault}{\updefault}{\color[rgb]{0,0,0}$s(\textbf{b},c,\zeta_1)$}%
}}}}
\end{picture}%
}
\end{minipage}\hspace{-5mm}
\begin{minipage}[t]{0.25\textwidth}\centering
\resizebox{!}{3.2cm}{\begin{picture}(0,0)%
\includegraphics{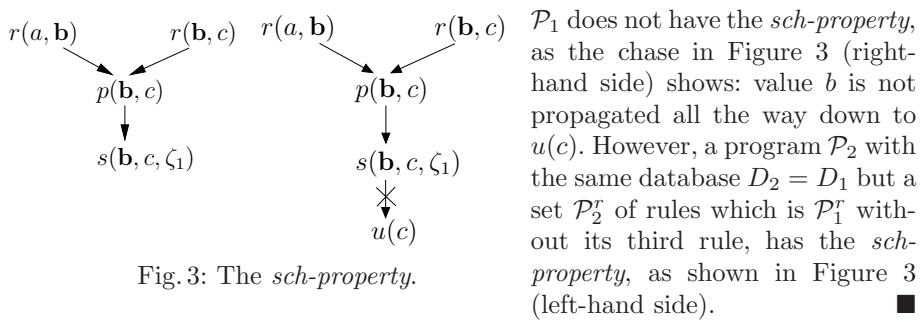}%
\end{picture}%
%
%
\setlength{\unitlength}{3947sp}%
\begingroup\makeatletter\ifx\SetFigFont\undefined%
\gdef\SetFigFont#1#2#3#4#5{%
  \reset@font\fontsize{#1}{#2pt}%
  \fontfamily{#3}\fontseries{#4}\fontshape{#5}%
  \selectfont}%
\fi\endgroup%
\begin{picture}(1480,1803)(458,-1248)
\put(473,367){\makebox(0,0)[lb]{\smash{{\SetFigFont{12}{14.4}{\rmdefault}{\mddefault}{\updefault}{\color[rgb]{0,0,0}$r(a,\textbf{b})$}%
}}}}
\put(1773,367){\makebox(0,0)[lb]{\smash{{\SetFigFont{12}{14.4}{\rmdefault}{\mddefault}{\updefault}{\color[rgb]{0,0,0}$r(\textbf{b},c)$}%
}}}}
\put(1182,-106){\makebox(0,0)[lb]{\smash{{\SetFigFont{12}{14.4}{\rmdefault}{\mddefault}{\updefault}{\color[rgb]{0,0,0}$p(\textbf{b},c)$}%
}}}}
\put(1182,-649){\makebox(0,0)[lb]{\smash{{\SetFigFont{12}{14.4}{\rmdefault}{\mddefault}{\updefault}{\color[rgb]{0,0,0}$s(\textbf{b},c,\zeta_1)$}%
}}}}
\put(1300,-1169){\makebox(0,0)[lb]{\smash{{\SetFigFont{12}{14.4}{\rmdefault}{\mddefault}{\updefault}{\color[rgb]{0,0,0}$u(c)$}%
}}}}
\end{picture}%
}
\end{minipage}
\vspace{-2mm}
\captionof{figure}{The {\em sch-property}.}\label{fig:chase}
\end{minipage}\hspace{2mm}
\begin{minipage}[t]{0.42\textwidth}
\vspace{-3.2cm}
\noindent $\mc{P}_1$ does not have the {\em sch-property}, as the chase in Figure~\ref{fig:chase} (right-hand side) shows: value $b$ is not propagated all the way down to $u(c)$. However, a program $\mc{P}_2$ with the same database $D_2=D_1$ but a set $\mc{P}_2^r$ of rules which is $\mc{P}_1^r$ without its third rule, has the {\em sch-property}, as shown in Figure~\ref{fig:chase} (left-hand side).\boxtheorem
\end{minipage}

\SCh \ is the semantic class of programs with the {\em sch-property}. Next, we briefly recall the classes of programs whose definitions are related to the {\em sch-property} and the \SCh \ programs.

\vspace{-5mm}
\subsubsection{Sticky Programs.} \label{sec:sticky} Sticky \dpm \ is a syntactic class of programs that enjoy the {\em sch-property}, for any extensional database \cite{cali12}.
Its programs are characterized through a body variable {\em marking procedure} whose input is the set $\mc{P}^r$ of program rules (the data do not participate).

The procedure has two steps: (a) {\em Preliminary step}, for each $\sigma \in \mc{P}^r$ and variable $X \in {\it body}(\sigma)$, if there is an atom $A \in {\it head}(\sigma)$ where $X$ does not appear, mark each occurrence of $X$ in ${\it body}(\sigma)$, and (b) {\em Propagation step}, for each $\sigma \in \mc{P}^r$, if a marked variable in ${\it body}(\sigma)$ appears at position $\pi$, then for every $\sigma' \in \mc{P}^r$ (including $\sigma$), mark each occurrence of the variables in ${\it body}(\sigma')$ that appear in ${\it head}(\sigma')$ in the same position $\pi$.

$\mc{P}^r$ is {\em sticky} when, after applying the marking procedure, there is no rule with a marked variable appearing more than once in its body (notice that a variable never appears both marked and unmarked in a  same body).

\vspace{-2mm}
\begin{example} \label{example:sticky}
The initial set of three rules, $\mc{P}^r$, is shown on the left-hand side below. The second rule already shows marked variables (with hat) after the preliminary step. The set of rules on the right-hand side are the result of whole marking procedure.\end{example}\vspace{-7mm}

\[ \arraycolsep=0pt
\begin{array}{rl c rl}
r(X,Y),p(X,Z) ~\rightarrow~& s(X,Y,Z). &\hspace{1cm}& r(\hat{X},Y),p(\hat{X},\hat{Z}) ~\rightarrow~& s(X,Y,Z).\\
s(\hat{X},Y,\hat{Z}) ~\rightarrow~& u(Y).&\hspace{1cm}&s(\hat{X},Y,\hat{Z}) ~\rightarrow~& u(Y).\\
u(X) ~\rightarrow~& \exists\;Y\;r(Y,X).&\hspace{1cm}&u(X) ~\rightarrow~& \exists\;Y\;r(Y,X).
\end{array}
\]
\vspace{-2mm}

Variables  $X$ and $Z$ in the first rule-body end up marked  after the propagation step: they appear in the same rule's head, in marked positions ($s[1]$ and $s[3]$ in the body of the second rule).
Accordingly, the set of rules is {\em not} sticky: $X$ in the first rule's body is marked and occurs twice (in $r[1]$ and $p[1]$).
\boxtheorem

With sticky programs,  QA can be done in polynomial-time in data complexity  \cite{cali12}. A program with the {\em sch-property} may not be syntactically sticky. Actually, the \SCh \ class can be extended to several larger, semantic, classes of programs that enjoy a form of the {\em sch-property} with the propagation condition during the chase only on values in certain forms of ``infinite" positions. (We propose a new, syntactic class along these lines   in Section~\ref{sec:pclasses}). Something similar can be done with the class of sticky programs.

\vspace{-5mm}
\subsubsection{Weakly-Sticky (\!\WS) Programs.} \label{sec:ws} This is a syntactic class that extends those of \WA \ and sticky programs.  Its characterization uses the above notions of finite-rank and marked variable: A set of rules $\mc{P}^r$ is \WS \ if, for every rule in it and every repeated variable in its body, the variable is either non-marked or appears in some position in $\Pi_F(\mc{P}^r)$.

\vspace{-2mm}
\begin{example} (example \ref{example:sticky} cont.) \ $\mc{P}^r$ is \WS, because $p[1] \in \Pi_F(\mc{P}^r)$; and $X$, the only repeated variable in a body (of the first rule),  is marked, but in $p[1]$.\boxtheorem
\end{example}
\vspace{-3mm}


The \WS \ condition guarantees tractability of QA, because CQs can be answered on an initial fragment of the chase whose size is polynomial  in that of the extensional database. This relies on these facts: (a) Finite-rank positions can be saturated by polynomially many values in the size of the extensional database. (b) Stickiness for infinite-rank positions ensures that polynomially many values are required in them for answering a query at hand. In fact, stickiness for infinite positions makes the number of values required in them for QA polynomially depend on the number of values in finite-rank positions. So, both in finite and infinite-rank positions, polynomially many values are needed.

The above argument about QA is more general than as applied to \WS \ programs. It can be applied with more general, syntactic and semantic, classes of programs that are characterized through the use of the stickiness condition on positions where infinitely many values may appear during the chase. \WS \ programs are a special case, where those positions are with infinite-rank; and the stickiness is enforced by the syntactic variable-marking mechanism. Actually, we can make the general claim that the combination of finitely many values in finite positions plus chase-stickiness on infinite positions makes QA decidable.

\vspace{-5mm}
\subsubsection{Generalized Stickiness.}The {\em generalized-stickiness of the chase (gsch-property)} is defined by relaxing the condition in the {\em sch-property}: the condition applies to values for the repeated body variables that do not appear in {\em finite positions}. \GSCh \ is the semantic class of programs with the {\em gsch-property} (cf. Figure~\ref{fig:range-classes}).

\vspace{-2mm}
\begin{example}(ex. \ref{exp:chs} cont.)\label{ex:gsch} $\mc{P}_1$ and $\mc{P}_2$ have no infinite positions because for both programs the chase terminates. Consequently, they are \GSCh. Consider a program $\mc{P}_3$ with the same database $D_3=D_1$ and a set $\mc{P}_3^r$ of rules which is $\mc{P}_2^r \cup \{\sigma\}$ such that, $\sigma\!: \ r(X,Y) \rightarrow \exists Z\;r(Z,X)$. $r[1]$ and $r[2]$ are infinite positions because, during the chase of $\mc{P}_3$, $\sigma$ cyclically generates infinite null values in $r[2]$ that also propagate to $r[1]$. The chase of $\mc{P}_3$ does not have the {\em gsch-property} and it is not \GSCh \ since the value $b$ replaces the repeated body variable $Y$ that only appears in infinite positions ($r[1]$ and $r[2]$) and $b$ does not propagate all the way down during the chase procedure.\boxtheorem\end{example}

\vspace{-8mm}
\section{Selection Functions and Program Classes}\label{sec:pclasses}
\vspace{-2mm}

The finite positions in the definition of the {\em gsch-property} are not computable for a given program which makes it impossible to decide if the program has the property. Here, we define selection functions that determine subsets of the finite positions of a program. We replace finite positions in the definition of the {\em gsch-property} with the results from selection functions in order to define new stickiness properties and program classes.

A {\em selection function} $\mc{S}$ (over a schema $\mc{R}$) is a function that takes a program $\mc{P}$ and returns a subset of $\fp{\mc{P}}$. Particular functions are $\mc{S}^\bot$ and $\mc{S}^\top$, that given a program $\mc{P}$, return the empty set and $\fp{\mc{P}}$, respectively. The latter may not be computable, and depends on the program's data, which is not the case for the former. $\Pi_F$ also defines a data-independent selection function, $\mc{S}^\nit{rank}$, that returns the finite-rank positions (there are finitely many values in them in the chase of $\mc{P}$, for any data set \cite[Lemma 5.1]{cali12}). A selection function is ``syntactically computable" if it only depends on the rules $\mc{P}^r$ of a program $\mc{P}$, and we use the notation $\mc{S}(\mc{P}^r)$.

The {\em $\mc{S}$-stickiness} is defined by replacing the finite positions in the definition of the {\em gsch-property} with a selection function $\mc{S}$: The chase of a program $\mc{P}$ has the {\em $\mc{S}$-stickiness property} if the stickiness condition applies only to values replacing the repeated body variables that do not appear in a position of $\mc{S}(\mc{P})$. \SCh$(\mc{S})$ \ is the semantic class of programs with the {\em $\mc{S}$-stickiness}. In particular, \SCh \ $=$ \SCh$(\mc{S}^\bot)$, \GSCh \ $=$ \SCh$(\mc{S}^\top)$. Also, \WSCh \ $=$ \SCh$(\mc{S}^\nit{rank})$ is the class of programs with {\em weak-stickiness of the chase}. \SCh$(\mc{S})$ specifies a range of semantic classes of programs starting with \SCh, ending with \GSCh, and with \WSCh \ in between.

\SCh$(\mc{S})$ grows monotonically with $\mc{S}$: For selection functions $\mc{S}_1$ and $\mc{S}_2$ over schema $\mc{R}$, if $\mc{S}_1 \subseteq \mc{S}_2$, then \SCh$(\mc{S}_1) \subseteq $ \SCh$(\mc{S}_2)$. Here, $\mc{S}_1 \subseteq \mc{S}_2$ if and only if for every program $\mc{P}$, $\mc{S}_1(\mc{P}) \subseteq \mc{S}_2(\mc{P})$. In general, the more finite positions are (correctly) identified (and the consequently, the less finite positions are treated as infinite), the more general subclass of \GSCh \ that is identified or characterized.

Sticky \dpm \ uses the marking procedure to restrict the repeated body variables and impose the {\em sch-property}. Applying this syntactic restriction only on body variables specified by syntactic selection functions results in syntactic classes that extend sticky \dpm. These syntactic classes are subsumed by the semantic classes defined by the same selection functions; each of these syntactic classes only partially represents its corresponding semantic class. Particularly, \SCh \ subsumes sticky \dpm~\cite{cali12}; and \WS \ is a syntactic subclass of \WSCh \ (cf. $(g)$ and $(h)$ in Figure~\ref{fig:range-classes}). \ignore{Example~\ref{ex:wsch} in Appendix~\ref{sec:examples} shows that the inclusion of \WS \ in \WSCh \ is proper which}

\vspace{-3mm}
\section{Joint-Weakly-Sticky Programs} \label{sec:jws}
\vspace{-3mm}

The definition of the class of \JWS \ programs uses the syntactic selection function $\mc{S}^\exists$, which appeals to the {\em existential dependency graph} of a program~\cite{krotzsch}
(to define {\em joint-acyclic} programs). We briefly review it here.

Let $\mc{P}^r$ be a set of rules that is standardized apart, i.e. no variable appears in more than one rule. For a variable $X$, let $B(X)$ ($H(X)$) be the set of all positions where $X$ occurs in the body (head) of its rule $\sigma$. For a $\exists$-variable $Z$, the set of target positions of $Z$, denoted by $T(Z)$, is the smallest set of positions such that (a) $H(Z) \subseteq T(Z)$, and (b) $H(X) \subseteq T(Z)$ for every $\forall$-variable $X$ with $B(X) \subseteq T(Z)$. Roughly speaking, $T(Z)$ is the set of positions where the null values invented by $Z$ may appear in during the chase.

An {\em existential dependency graph} (\edg) of $\mc{P}^r$ is a directed graph with the $\exists$-variables of $\mc{P}^r$ as its nodes. There is an edge from $Z$ to $Z'$ if there exists a body variable $X$ in the rule containing $Z'$ such that $B(X) \subseteq T(Z)$. Intuitively, the edge shows that the values invented by $Z$ might appear in the body of the rule of $Z'$ and cause invention of values by $Z'$. Therefore, a cycle represents the possibility of infinite null values invention by the $\exists$-variables in the cycle.

\ignore{A program is {\em joint-acyclic} (\JA) if its \edg \ is acyclic. \JA \ programs have polynomial size chase w.r.t the size of the extensional data and properly extend \WA \ programs~\cite{krotzsch}.}

\vspace{-2mm}
\begin{example} \label{ex:edg}  Let $\mc{P}^r$ contain the following rules: $u(Y),r(X,Y) \rightarrow \exists Z\;r(Y,Z)$ and $r(X',Y'),r(Y',Z') \rightarrow p(X',Z')$. For the variable $Y$,
$B(Y)=\{u[1],r[2]\}$, $H(Y)=\{r[1]\}$. Moreover, $T(Z)=\{r[2],p[2]\}$. The \edg \ of $\mc{P}^r$ has $Z$ as its node without any edge since $B(X)$ and $B(Y)$ are not subsets of $T(Z)$.
\ignore{The graph is acyclic, so $\mc{P}^r$ is \JA. Notice that} $\mc{P}^r$ is not \WA, because $r[1]$ and $r[2]$  have infinite rank.\boxtheorem\end{example}\vspace{-4mm}


\ignore{\begin{figure}
\begin{center}
\includegraphics[width=11cm]{generalizations.eps}
\end{center}
\vspace{-0.3cm}
\caption{Generalization relationships between program classes}
\label{fig:gener}
\end{figure}}




For a set of rules $\mc{P}^r$, we define the set of {\em finite-existential positions} of $\mc{P}^r$ denoted by $\Pi^\exists_F(\mc{P}^r)$ as follows: It is the set of positions that are not in the target set of any $\exists$-variable in a cycle in \edg$(\mc{P}^r)$. Intuitively, a position in $\Pi^\exists_F(\mc{P}^r)$ is not in the target of any $\exists$-variable that may invent infinite null values.

\vspace{-2mm}
\begin{proposition} \em \label{pr:selection} For every set of rules $\mc{P}^r$, $\Pi_F(\mc{P}^r)\subseteq \Pi^\exists_F(\mc{P}^r)$.\boxtheorem \end{proposition}\vspace{-4mm}

$\Pi^\exists_F$ defines a computable selection function $\mc{S}^\exists$ that returns finite-existential positions of a program (cf. $(c)$ in Figure~\ref{fig:range-classes}). \SCh$(\mc{S}^\exists)$ is a new semantic subclass of \GSCh \ that generalizes \SCh$(\mc{S}^\nit{rank})$ since $\mc{S}^\exists$ provides a finer mechanism for capturing finite positions in comparison with $\mc{S}^\nit{rank}$ (cf. $(e)$ and $(f)$ in Figure~\ref{fig:range-classes}).

A program $\mc{P}$ is {\em joint-weakly-sticky} (\JWS) \ if for every rule in $\mc{P}^r$ and every variable in its body that occurs more than once, the variable is either non-marked or appears in some positions in $\Pi^\exists_F(\mc{P}^r)$. The class of \JWS \ programs is a proper subset of \SCh$(\mc{S}^\exists)$ and extends \WS \ (cf. (i) and (k) in Figure~\ref{fig:range-classes}). Specifically, the program  in Example~\ref{ex:edg} is \JWS,  because every position is finite-existential, but not \WS, because $Y'$ is marked and appears in $r[1]$ and $r[2]$ with infinite rank.

\vspace{-5mm}
\section{A Chase-Based Query Answering Algorithm}\label{sec:qa}
\vspace{-2mm}

\QAA \ is a QA algorithm for programs in the semantic class of \SCh$(\mc{S})$. It is based on a bottom-up data generation approach and applies a query-driven chase. The algorithm takes as input a computable selection function $\mc{S}$, a program $\mc{P} \in $ \SCh$(\mc{S})$, and a CQ $\mc{Q}$ over schema $\mc{R}$ and returns $\nit{ans}(\mc{Q},\mc{P})$.

Before describing \QAA, we introduce some notations. A {\em homomorphism} is a structure-preserving mapping, $h\!\!:\mc{C}\cup\mc{N} \!\rightarrow \!\mc{C}\cup\mc{N}$, between two instances over schema $\mc{R}$ that is the identity on constants. An {\em isomorphism} is a bijective homomorphism.

\vspace{-2mm}
\begin{definition}\label{df:app}\em A rule $\sigma \in \mc{P}^r$ and an assignment $\theta$ are {\em applicable} over an instance $I$ of $\mc{R}$ if: \ (a) $I \models (\nit{body}(\sigma))[\theta]$; and (b) there is an assignment $\theta'$ that extends $\theta$, maps the $\exists$-variables of $\sigma$ into fresh nulls, and $\theta'(\nit{head}(\sigma))$ is {\em not isomorphic} to any atom in $I$.\boxtheorem\end{definition}
\vspace{-3mm}

Note that for an instance $I$ and a set of rules $\mc{P}^r$, we can systematically compute the applicable pairs of rule-assignment by first finding $\sigma \in \mc{P}^r$ for which $\nit{body}(\sigma)$ is satisfied by $I$. That gives an assignment $\theta$ for which $(\nit{body}(\sigma))[\theta] \in I$. Then, we construct $\theta'$ as specified in Definition~\ref{df:app} and we iterate over atoms in $I$ and we check if they are isomorphic to $\theta'(\nit{head}(\sigma))$.

In \QAA, we use the notion of {\em freezing a null value} that is moving it from $\mc{N}$ into $\mc{C}$. It may cause new applicable rule-assignment because it changes isomorphic atoms. Considering an instance $I$, the {\em resumption} of a step of \QAA \ is freezing every null in $I$ and continuing the step. Notice that a pair of rule-assignment is applied only once in Step~2. Moreover, if there are more than one applicable pairs, then \QAA \ chooses the pair that becomes applicable sooner. \QAA \ is applicable to any \dplus \ program and any selection function, and returns sound answers. However, completeness is guaranteed only when applied to programs in \SCh$(\mc{S})$ with a computable $\mc{S}$.

\begin{algorithm}[h]
\caption{The \QAA \ algorithm}
\begin{algorithmic}[1,h]
\Statex{{\bf Inputs:} A selection function $\mc{S}$, a program $\mc{P} \in $ \SCh$(\mc{S})$, and a CQ $\mc{Q}$ over $\mc{P}$.}
\Statex{{\bf Output:} $\nit{ans}(\mc{Q},\mc{P})$.}
\Statex{}
\Statex{{\bf Step~1:} Initialize an instance $I$ with the extensional database $D$.}\vspace{1mm}
\Statex{{\bf Step~2:} Choose an applicable rule-assignment $\sigma$ and $\theta$ over $I$, add ${\it head}(\sigma)[\theta']$ into $I$ in which $\theta'$ is an extension of $\theta$ with mappings for the $\exists$-variables in $\sigma$ to fresh nulls in $\mc{N}$.}\vspace{1mm}
\Statex{{\bf Step~3:} Freeze the nulls in the new atom in Step~2 that appear in the positions of $\mc{S}(\mc{P})$.}\vspace{1mm}
\Statex{{\bf Step~4:} Iteratively apply Steps~2 and 3 until there is no more applicable pair of rule-assignment.} \vspace{1mm}
\Statex{{\bf Step~5:} Resume Step~2 with $I$, i.e. freeze nulls in $I$ and continue with Steps~2. Repeat resumption $M_\mc{Q}$ times where $M_\mc{Q}$ is the number of variables in $\mc{Q}$} \vspace{1mm}
\Statex{{\bf Step~6:} Return the tuples in $\mc{Q}(I)$ that do not have null values (including the frozen nulls).}
\end{algorithmic}
\end{algorithm}

\vspace{-2mm}
\begin{example} Consider a program $\mc{P}$ with $D= \{s(a,b,c), $ \ $v(b), u(c) \}$, and a BCQ $\mc{Q}: p(c,Y)\rightarrow \nit{ans}_\mc{Q}$, and a set of rules $\mc{P}^r$ containing (the hat signs show the marked variables):

\vspace{-6mm}
\[ \arraycolsep=0pt
\begin{array}{r rl c r rl}
\sigma_1:&\hspace{2mm}s(\hat{X},\hat{Y},\hat{Z})&\rightarrow\exists W\;s(Y,Z,W). &\hspace{1cm}&\sigma_2:&\hspace{2mm}u(\hat{X})&\rightarrow\exists Y,Z\;s(X,Y,Z).
\end{array}
\]
\vspace{-7mm}
\[ \arraycolsep=0pt\def\arraystretch{1.8}
\begin{array}{r rl}
\sigma_3:& ~s(\hat{X},Y,Z),v(\hat{X}),s(Y,Z,\hat{W})&\rightarrow p(Y,Z).
\end{array}
\]

\vspace{-2mm}
$\mc{P}$ is in \WS \ and so \SCh$(\mc{S}^\nit{rank})$. Specifically in $\sigma_3$, $X$ occurs in $v[1]$ which is in $\mc{S}^\nit{rank}(\mc{P}^r)$ and $Y$ and $Z$ are not marked. The algorithm starts from $I=D$. At Step~2, $\sigma_1$ and $\theta_1\!=\!\{X\!\!\rightarrow\! a,Y\!\!\rightarrow \!b,Z\!\!\rightarrow \!c\}$ are applicable; and \QAA \ adds $s(b,c,\zeta_1)$ into $I$. $\sigma_2$ and $\theta_2\!=\!\{X\!\rightarrow\!c\}$ are also applicable and they add $s(c,\zeta_2,\zeta_3)$ into $I$. Note that Step~3 does not freeze $\zeta_1$, $\zeta_2$, and $\zeta_3$ since they are not in $\mc{S}^\nit{rank}(\mc{P}^r)$

There is not more applicable rule-assignments and we continue with Step~5. Consider that $\sigma_1$ and $\theta_3\!=\!\{X\!\!\rightarrow \!b,Y\!\!\rightarrow \!c,Y\!\!\rightarrow \!\zeta_1\}$ are not applicable since any $\theta'_3\!=\!\theta_3\cup \{W \!\!\rightarrow \!\zeta_4\}$ generates $s(c,\zeta_1,\zeta_4)$ that is isomorphic with $s(c,\zeta_2,\zeta_3)$ already in $I$. \QAA \ is resumed once since $\mc{Q}$ has one variable. This is done by freezing $\zeta_1,\zeta_2,\zeta_3$ and returning to Step~2. Now, $s(c,\zeta_1,\zeta_4)$ and $s(c,\zeta_2,\zeta_3)$ are not isomorphic anymore and $\sigma_1$ and $\theta_3$ are applied which results in $s(c,\zeta_1,\zeta_4)$. As a consequence, $\sigma_3$ and $\theta_4=\{X\rightarrow b,Y\rightarrow c,Z\rightarrow \zeta_1,W\rightarrow \zeta_4\}$ are applicable, which generate $p(c,\zeta_1)$. The instance $I$ in Step~6 is $I=D\cup\{s(b,c,\zeta_1),s(c,\zeta_2,\zeta_3),s(c,\zeta_1,\zeta_4),p(c,$ \ $\zeta_1),s(\zeta_2,\zeta_3,\zeta_5),s(\zeta_1,\zeta_4,\zeta_6)\}$, and $I \models \mc{Q}$.\boxtheorem \end{example}
\vspace{-3mm}

The number of resumptions with \QAA \ depends on the query. However, for practical purposes, we could run \QAA \ with $N$~resumptions,  to be able to answer queries with up to ~$N$ variables. If a query has more than $N$ variables, we can incrementally retake the already-computed instance $I$, adding the required number of resumptions.

\vspace{-2mm}
\begin{theorem}\label{th:correctness} \em Consider a computable selection function $\mc{S}$, a program $\mc{P} \in $ \SCh$(\mc{S})$, and a CQ $\mc{Q}$ over schema $\mc{R}$. Algorithm \QAA \ taking $\mc{S}$, $\mc{P}$, and $\mc{Q}$ as inputs, terminates returning $\nit{ans}(\mc{Q},\mc{P})$.\boxtheorem \end{theorem}
\vspace{-4mm}

Termination is due to condition (b) in Definition~\ref{df:app}, which prevents isomorphic atoms in $I$. Note that because of Step~3 the null values that appear in the positions of $\mc{S}(\mc{P})$ are treated as constants while deciding isomorphic atoms. However, condition (b) in Definition~\ref{df:app} prevents some atoms from $I$ that are necessary for answering $\mc{Q}$. Adding these atoms depends on the applicability of certain pairs of rule-assignment in which the assignment replaces some repeated variables in the body of the rule with null values. Each resumption makes some of these pairs applicable by freezing nulls. Since $\mc{P}$ is \SCh$(\mc{S})$, there are at most $M_\mc{Q}$ such rules and so $M_\mc{Q}$ resumptions are sufficient for answering $\mc{Q}$. The running time of \QAA \ depends on the number of finite values that may appear in the positions of $\mc{S}(\mc{P})$.

\vspace{-3mm}

\begin{proposition}\label{pr:ptime} \em Algorithm \QAA \ runs in {\sc ptime} in data  if the following  holds for $\mc{S}$: for any program $\mc{P}'$, the number of values appearing in $\mc{S}(\mc{P}')$-positions during the chase is polynomial in the size of the extensional data.\boxtheorem \end{proposition}

\vspace{-7mm}

\begin{lemma}\label{lm:ptime}\em During the chase of a \dplus \ program $\mc{P}$, the number of distinct values in $\mc{S}^\exists(\mc{P}^r)$-positions is polynomial in the size of the extensional data.\boxtheorem\end{lemma}

\vspace{-7mm}

\begin{corollary} \em \QAA \ runs in {\sc ptime} in data with  programs in \SCh$(\mc{S}^\exists)$, in particular for the programs in the \JWS \ and \WS \ syntactic classes.\boxtheorem
\end{corollary}

\vspace{-6mm}

\section{Magic-Sets and \JWS \ \dpm}\label{sec:magic}
\vspace{-2mm}

Magic-sets is a general technique for rewriting logical rules so that they may be implemented bottom-up in a way that avoids the generation of irrelevant facts~\cite{beeri-ms,ceri}. The advantage of such a rewriting technique is that, by working bottom-up, we can take advantage of the structure of the query and the data values in it,  optimizing the data generation process.

In this section, we present a magic-sets rewriting for \dplus \ programs, denoted by \ms. It has two changes regarding the technique in~\cite{\ignore{bancilhon,beeri-ms,}ceri} in order to: (a) work with $\exists$-variables in the existential rules, and (b) consider the extensional data of the predicates that also have intensional data defined by the rules. For (a), we apply the solution proposed in~\cite{alviano}. However (b) is specifically relevant for \dplus \ programs that allow predicates with both extensional and intentional data, and we address it in \ms. \ms \ is described in detail in Appendix D in \cite{qaMS}.

\vspace{-2mm}
\begin{example} (ex. \ref{ex:edg} cont.) \label{ex:mg} Consider a BCQ $\mc{Q}:p(a,Y)~\rightarrow~\nit{ans}_\mc{Q}$ over a program $\mc{P}$ with $D=\{u(a),r(a,b)\}$ and the rules in $\mc{P}^r$. \ms \ has the following steps:

\vspace{-2mm}
\begin{enumerate}
\item Generate the adorned version of the query by annotating its body predicates with strings of $b$s and $f$s that correspond to the positions with constants or variables respectively. Then, propagate the adorned predicates to the other program rules. Here, $p^\nit{bf}(a,Y)~\rightarrow~\nit{ans}_\mc{Q}$ is the adorned query; $r^\nit{bf}(X,Y),r^\nit{bf}(Y,Z) ~\rightarrow~ p^\nit{bf}(X,Z)$ and $u(Y),r^\nit{fb}(X,Y) ~\rightarrow~ \exists Z\;r^\nit{bf}(Y,Z)$ are the adorned rules. Note that the first rule in $\mc{P}^r$ is not adorned by bounding $Z$ in the head (e.g. $r^\nit{fb}(Y,Z)$) since the $\exists$-variables can not be bounded.

\item  Add magic predicates to the body of the adorned rule. The magic predicates specify the values for the bounded variables: $\nit{mg}\_p^\nit{bf}(X),r^\nit{bf}(X,Y),r^\nit{bf}(Y,Z)$ $\rightarrow p^\nit{bf}(X,Z)$ and $\nit{mg}\_r^\nit{bf}(Y),u(Y),r^\nit{fb}(X,Y) \rightarrow \exists Z\;r^\nit{bf}(Y,Z)$.

\item  Generate magic rules that define the magic predicates: $\nit{mg}\_p^\nit{bf}(X) \rightarrow \nit{mg}\_r^\nit{bf}$ \ $(X)$ and $\nit{mg}\_r^\nit{bf}(X),r^\nit{bf}(X,Y) ~\rightarrow$ $\nit{mg}\_r^\nit{bf}(Y)$, and a fact $\nit{mg}\_p^\nit{bf}(a)$.

\item  For the adorned predicates with extensional data (e.g. $r$), generate new rules to load their extensional data: $\nit{mg}\_r^\nit{bf}(X),r(X,Y) ~\rightarrow~ r^\nit{bf}(X,Y)$ and $\nit{mg}\_r^\nit{fb}(Y),r(X,Y)~\rightarrow~ r^\nit{fb}(X,Y)$.
\end{enumerate}

\vspace{-2mm}
\noindent The result is a program $\mc{P}_m$ with schema $\mc{R}_m$, $D_m=D$, the set of rules $\mc{P}^r_m$ specified in Steps~2-5, and $\mc{Q}_m$ which is the adorned query from Step~1. \boxtheorem\end{example}

\vspace{-4mm}
\ms \ differs from the rewriting algorithm of~\cite{alviano} in Step~4. Particularly, in the latter Step~4 is not needed since, unlike the former, it assumes the intentional predicates in $\mc{P}$ and the adorned predicates in $\mc{P}_m$ do not have extensional data. Therefore, the correctness of \ms, i.e. $\nit{ans}(\mc{Q},\mc{P})=\nit{ans}(\mc{Q}_m,\mc{P}_m)$, follows from both the correctness of the rewriting algorithm in~\cite{alviano} and Step~4.

$\mc{P}^r_m$ has certain syntactic properties. First, the magic rules do not have $\exists$-variables. Also as mentioned in Step~1, the positions of $\exists$-variables in the head of a rule never become bounded. Additionally we assume that the full information about bounded variables is propagated from the head of an atom to its body. That is when a variable is in a bounded position in the head it appears in the body only in bounded positions. 

Applying \ms \ over a \WS \ program $\mc{P}$, $\mc{P}_m$ is not necessarily \WS \ or in \SCh$(\mc{S}^\nit{rank})$ (cf. Example~14  in Appendix~E in \cite{qaMS}), which means \SCh$(\mc{S}^\nit{rank})$ and \WS \ are not closed under \ms. This is because \ms \ introduces new join variables between the magic predicates and the adorned predicates, and these variables might be marked and appear only in the infinite rank positions. That means the joins may break the $\mc{S}^\nit{rank}$-stickiness as it happens in Example~14 in Appendix~E~\cite{qaMS}. Specifically it turned out to be because $\mc{S}^\nit{rank}$ decides some finite positions of $\mc{P}^r_m$ as infinite rank positions. In fact, the positions of the new join variables are always bounded and are finite. Therefore, \ms \ does not break $\mc{S}$-stickiness if we consider a finer selection function $\mc{S}$ that decides the bounded positions as finite. We show in Theorem~\ref{th:closed} that the class of \SCh$(\mc{S}^\exists)$ and its subclass of \JWS \ are closed under \ms \ since they apply $\mc{S}^\exists$ that better specifies finite positions compared to $\mc{S}^\nit{rank}$.

\vspace{-2mm}
\begin{theorem}\label{th:closed} \em Let $\mc{P}$ and $\mc{P}_m$ be the input and the result programs of \ms \ respectively. If $\mc{P}$ is \JWS, then $\mc{P}_m$ is \JWS.\boxtheorem \end{theorem}
\vspace{-4mm}

As a result of Theorem~\ref{th:closed}, we are able to apply \ms \ in order to optimize \QAA \ for the class of \JWS \ and its subclasses sticky and \WS.

\vspace{-3mm}
\section{Conclusion and Future Research}\label{sec:conc}
\vspace{-3mm}

We introduced semantic and syntactic extensions of sticky and \WS \ \dpm \ and we proposed a practical bottom-up QA algorithm for these programs. We applied a magic-set rewriting technique, \ms, to optimize the QA algorithm. As the future work, we intend to study the applications of the magic-set rewriting for \dpm \ ontologies and in the presence of program constraints, i.e. negative constraints and equality generating dependencies and specifically for the purpose of managing inconsistency for these ontologies. We believe that \QAA \ and \ms \ are applicable on real-world scenarios and we plan to implement them and run experiments on real-world data with large data sets.

\appendix

\newpage

\section{Proofs}\label{sec:proofs}

\defproof{Proposition~\ref{pr:selection}}{We use proof by contradiction. Assume there is a position $\pi$ such that: $\pi \in \Pi_F(\mc{P}^r)$ and $\pi \not\in \Pi^\exists_F(\mc{P}^r)$. The latter means there is a cycle in \edg$(\mc{P}^r)$ that includes an $\exists$-variable $Z$ in a rule $\sigma$ such that $\pi \in T(Z)$. The definition of \edg \ implies that, there is $\forall$-variable $X$ in the body of $\sigma$ for which $B(X) \subseteq T(Z)$. Let $\pi_Z$ and $\pi_X$ be the two positions where $Z$ and $X$ appear in $\sigma$ resp. Then, there is a path from $\pi_Z$ to $\pi_X$ and there is also a special edge from $\pi_X$ to $\pi_Z$ in \dg$(\mc{P}^r)$ making a cycle including $\pi_Z$ with a special edge. Therefore, $\pi_Z \not\in \Pi_F(\mc{P}^r)$. Since $\pi \in T(Z)$, we can conclude that $\pi \not\in \Pi_F(\mc{P}^r)$ which contradicts the assumption and completes the proof.}

\defproof{Theorem~\ref{th:correctness}}{Let $I_i$ be the instance $I$ after the $i$-th resumption in \QAA. To prove the termination, we first show that for a finite $i$ there are finitely many terms and so finitely many atoms in $I_i$. Since the algorithm only adds atoms this suffices to prove the algorithm always stops by reaching a fixed point.

Now, let $f^\mc{P}$ be the number of terms (constants and nulls) that appear in the positions of $\mc{S}(\mc{P})$ in $I$ during \QAA, and let $r^\mc{P}$ and $w^\mc{P}$ be the number of distinct predicate names and the maximum arity of predicates in $\mc{P}$ respectively. Starting from $I_0$, since there are no isomorphic atoms in $I_0$, there are at most $r^\mc{P} \times w^\mc{P}$ nulls (not frozen) and $r^\mc{P}\times w^\mc{P} + f^\mc{P}$ possible terms it $I_0$. Considering $r^\mc{P},w^\mc{P},f^\mc{P}$ are finite, $I_0$ is finite. After the first resumption, the $r^\mc{P}\times w^\mc{P}$ nulls are frozen; and at most $r^\mc{P} \times w^\mc{P}$ new nulls are invented. Now in $I_1$, there are at most $2\times r^\mc{P}\times w^\mc{P} + f^\mc{P}$ terms which means $I_1$ is also finite. With the same line of reasoning, we can prove that $I_i$ with a finite $i$ has finite terms, $i\times r^\mc{P}\times w^\mc{P} + f^\mc{P}$, and it is finite. Since there are $M_\mc{Q}$ resumptions and $M_\mc{Q}$ is finite, \QAA \ terminates.

\QAA \ is sound because Step~2 is sound and it only adds atoms into $I$ that are entailed by the rules in $\mc{P}^r$.

For the proof of completeness, we assume $\mc{Q}$ is a BCQ. Note that for free CQs we can make a BCQ for every tuple in the answers set and apply the same proof for the obtained BCQs. To prove the completeness of \QAA, i.e. $\mc{P} \models \mc{Q} \Rightarrow I_{M_\mc{Q}} \models \mc{Q}$, it is enough to show $I_\infty \models \mc{Q} \Rightarrow I_{M_\mc{Q}} \models \mc{Q}$. That is because $I_\infty$ (the instance after infinitely many resumptions) gives the same answers that are obtained from the chase of $\mc{P}$, since every null value in $I_\infty$ is eventually frozen and condition (b) in Definition~\ref{df:app} is always satisfied.

Let $\mc{P} \models \mc{Q}$ then, as it is proved in~\cite{cali12}, there is a {\em proof-schema} $T$ for $Q$ w.r.t $\mc{P}$. A proof-schema (called {\em accepting resolution proof-schema} in~\cite{cali12}) is a tree with its nodes and edges labeled with atoms of the schema $\mc{R}\cup \{\nit{ans}_\mc{Q}\}$ and the rules in $\mc{P}^r\cup\{\mc{Q}\}$ resp. The terms in the atoms are either constants in $D$ or variables. In $T$, the root node is labeled with $\nit{ans}_\mc{Q}$ and there is an assignment $\theta$ of the variables in the labels of the nodes in $T$ into the constants in $D$ and nulls that maps the labels of the nodes (other than the root node) into the atoms in $\nit{chase}(\mc{P})$. For every leaf node, $h$ maps its label into an atom in $D$. The label of the incoming edges into a node are a rule that shows how the atom of the node is obtained from the atoms in its child nodes. A proof-schema has other syntactic properties that are described in \cite[Definition~3.5]{cali12}. Without loss of generality, we assume that (a) $T$ has minimum height, and (b) $\theta$ maps $T$ into the atoms of $\nit{chase}(\mc{P})$ that are obtained sooner during the chase procedure. In the rest of the proof, whenever we refer to a node as an atom we mean the atom in the label of the node.

Since $I_\infty \models \mc{Q}$, there is also an assignment $\theta'$ that maps nodes of $T$ into the atoms in $I_\infty$.  The rest of the proof is devoted to show that $\theta'$ maps the nodes of $T$ into $I_{M_\mc{Q}}$ which proves $I_{M_\mc{Q}} \models \mc{Q}$. We do that by showing every variable in $T$ that appears in more than one branch is mapped by $\theta'$ into a term that is either a constant or a frozen null in $I_{M_\mc{Q}}$.

Let $X_1,...,X_n$ be the variables that appear in more than one branch of $T$ and do not occur in any position of $\mc{S}(\mc{P})$, ordered by the depth they first occur in $T$ ($X_1$ is the deepest). The $\mc{S}$-stickiness implies that $n \le M_\mc{Q}$. That is because these variables represent joins between values that do not appear in the $\mc{S}(\mc{P})$ positions and so they propagate all the way to the query. Therefore the number of these values (and so the variables) is restricted by the number of variables in the query. Also, let $A_1,...,A_n$ be the nodes (atoms) in $T$ where $X_1,...,X_n$ first appear. We claim that $\theta'(A_i)$ is in $I_{i-1}$ for each $i \le n$.

Consider $T_1$, the subtree of $A_1$. Its leaf nodes are mapped by $\theta'$ into $D\subseteq I_0$ according to the definition of $T$. In the internal nodes, if a variable appears in more than one branch of $T_1$, it occurs at least once in a position of $\mc{S}(\mc{P})$ in each branch. Now consider the variable $Y$ as the first such variable and $A_Y$ as the node where it first appears in $T_1$ and $A'_Y$ as the node where the branches meet. $A_Y$ is in $I_0$ because of the assumptions (a) and (b). Additionally, if the term $t=h'(Y)$ is a null value, it is frozen in $I_0$ because it appears in some positions of $\mc{S}(\mc{P})$. Note that even if $Y$ occurs in $A_Y$ in a non-$\mc{S}(\mc{P})$ position which means $t$ is not frozen immediately in Step~3 of \QAA, $t$ will eventually becomes frozen in $I_0$ before reaching $A'_Y$. That is because any other isomorphic atom $B$ with the term $t'$ that prevented $A_Y$ from $I_0$ (according to condition (b) in Definition~\ref{df:app}) will eventually propagate to the same position $\pi$ and becomes frozen and will not be isomorphic to $A_Y$ anymore. Note that we assumed there is no other join, so if $t$ was going to propagate to $A'_Y$, $t'$ will also propagate to $A'_Y$ and its $\mc{S}$-finite position. Similarly, we can prove that every variable that appears in more than one branch of $T_1$ is mapped by $\theta'$ into a term that is either constant or is frozen in $I_0$. Therefore every term in the atoms of the subtree $T_1$ are frozen in $I_0$ and so the nodes in $T_1$ are mapped by $\theta'$ into $I_0$.

Now since $\theta'(A_1)$ is in $I_0$, the term $\theta'(A_1)$ is frozen in $I_1$. Similarly, we can prove that $A_2$ is in $I_1$ considering that $\theta'(A_1)$ is frozen in $I_1$ and continuing with this line of reasoning we can prove that $\theta'(A_i)$ is in $I_{i-1}$. That means every join variable in $T$ is mapped by $\theta'$ into either a constant or a null that is frozen in $I_n$ with $n \le M_\mc{Q}$. Therefore $T$ is mapped by $\theta'$ into $I_{M_\mc{Q}}$ which completes our proof of the completeness of \QAA.}

\defproof{Proposition~\ref{pr:ptime}}{The condition implies that $f^\mc{P}$ (cf. the proof of Theorem~\ref{th:correctness}) is polynomial w.r.t the extensional data of $\mc{P}$. As a result, the number of terms in $I_i$ (the instance in \QAA \ after $i$-th resumption) $i\times r^\mc{P}\times w^\mc{P} + f^\mc{P}$ and also the size of $I_i$ are polynomial in the size of the extensional data. Since the algorithm only adds atoms to the current instance $I$ (never removes atoms from $I$), that means \QAA \ stops in {\sc ptime} in the size of extensional data.}

\defproof{Lemma~\ref{lm:ptime}}{The proof is similar to the proof of~\cite[Theorem 3.9]{fagin}. The theorem shows the chase of a \WA \ program has polynomial length in the size of the extensional data of the program.

We define $\exists$-rank of a position $\pi$ in a predicate in $\mc{P}^r$ as the maximum length of a path in \edg$(\mc{P})$ ending with $Z$ such that $\pi \in T(Z)$. A finite-existential position has a finite $\exists$-rank, since it is not in the target of any $\exists$-variable that is in a cycle in \edg$(\mc{P})$.

We prove by induction that: For every finite $i > 0$, there is a polynomial function $f_i$ such that the number of values that appear in the positions with $\exists$-rank $i$ is at most $f_i(d)$ with $d=\nit{size}(D)$.

{\em Base case}: The positions with $\exists$-rank of 0 are not in the target of any $\exists$-variable. Therefore, these positions can only contain constants from $D$, and $f_0=d$.

{\em Inductive step}: The values that appear in a position of $\exists$-rank $i$ are either (a) from the other positions with the same $\exists$-rank, or (b) from positions with the $\exists$-rank $j < i$. For (b), they are by inductive hypothesis at most $f_{i-1}(d)$. In case of (a), the values are invented by an $\exists$-variable $Z$ that is at the end of a path of length $i$ in \edg$(\mc{P})$. If there are $b_Z$ variables in the body of the rule of $Z$, the rule can invent $f_{i-1}(d)^{b_Z}$ new values for the positions with $\exists$-rank $i$. There are at most $s_\mc{P}$ such $\exists$-variables where $s_\mc{P}$ is the maximum number of rules in $\mc{P}^r$. Therefore $f_i(d)=s_\mc{P}\times f_{i-1}(d)^{b_Z}+f_{i-1}(d)$ and since $s_\mc{P}$ and $b_Z$ are independent of data, $f_{i}$ is {\sc ptime} w.r.t $d$.

Considering that $i \le k$ and $k$ (the maximum $\exists$-rank in $\mc{P}$) is independent of the data of $\mc{P}$, we conclude that $f_k(d)$ is the polynomial maximum number of distinct values in the positions of $\Pi_F^\exists(\mc{P}^r)$ which proves the proposition.}

\defproof{Theorem~\ref{th:closed}}{To prove $\mc{P}_m$ is in \SCh$(\mc{S}^\exists)$ we show every repeated variable in $\mc{P}_m$ preserves the $\mc{S}^\exists$-stickiness property.

First we claim that every bounded position in $\mc{P}_m$ is in $\Pi_F^\exists(\mc{P}_m)$. That is specifically because an $\exists$-variable never gets bounded during \ms \ and also if a position in the head is bounded the corresponding variable appears in the body only in the bounded positions. As a result, a bounded position can not be in the target of any $\exists$-variable which proves the claim.

Also note that if a position in $\mc{P}$ is finite-existential (the position is in $\Pi_F^\exists(\mc{P})$), its corresponding position in $\mc{P}_m$ is also finite-existential. The prove is by assuming that there is a finite-existential position $\pi \in \mc{P}$ and its corresponding position $\pi' \in \mc{P}_m$ is not finite-existential which means there is a loop in the \edg \ of $\mc{P}_m$ including a variable $Z'$ such that $\pi' \in T(Z')$. Then it is easy to show there is also a loop in the \edg \ of $\mc{P}$ including a variable $Z$ and $\pi \in T(Z)$ meaning that $\pi$ is not finite-existential which contradicts the assumption and completes the proof.

Now, we specify four types of joins in $\mc{P}_m$: (a) between the adorned predicates in the adorned rules, (b) between the adorned predicates in the magic rules, (c) between the adorned predicates and the magic predicates in the adorned rules, and (d) between the adorned predicates and the magic predicates in the magic rules.

The joins of Type~(a) do not break the $\mc{S}^\exists$-stickiness property since they correspond to join variables in $\mc{P}$. If they were not marked in $\mc{P}$ they are still not marked in $\mc{P}_m$ and if they were at some finite-existential position the same holds for the variable in $\mc{P}_m$ and either way the repeated variable in $\mc{P}_m$ preserves the $\mc{S}^\exists$-stickiness property. The joins of Type~(b), (c), and (d) also preserve the property since their variables appear in a bounded position and we proved the bounded positions are finite-existential. Therefore every type of joins in $\mc{P}_m$ satisfies the $\mc{S}^\exists$-stickiness property and so $\mc{P}_m$ is in \SCh$(\mc{S}^\exists)$.

Note that the same prove holds for \JWS \ programs, while it does not apply to \WS \ and \SCh$(\mc{S}^\nit{rank})$. The latter because the two claims at the beginning of the proof does not hold for these programs.}

\section{The Chase Procedure}\label{sec:chase}

The chase procedure of a program $\mc{P}$ with database $D$ and rules $\mc{P}^r$ starts from the extensional database $D$ and it iteratively applies the rules in $\mc{P}^r$ through some chase steps. In a chase step, the procedure applies a rule $\sigma \in \mc{P}^r$ and an assignment $\theta$ on the current instance $I$. $\sigma$ and $\theta$ are applicable if $\theta$ maps the body of $\sigma$ into $I$. Let $\theta'$ be an extension of $\theta$ that maps the $\exists$-variables of $\sigma$ into fresh nulls in $\mc{N}$. The result of applying $\sigma$ and $\theta$ over $I$ is an instance $I'=I\cup \{\theta'({\it head}(\sigma))\}$. We denote a chase step by $I \xrightarrow{\sigma,\theta} I'$.

Based on chase steps, the {\em level} of an atom is defined as follows: For an atom $a \in D$, $\nit{level}(a) = 0$. If an atom is the result of a chase step, $I_{i-1}\xrightarrow{\sigma_i,\theta_i}I_i$, let $\nit{level}(a)\!=\!\max_{\{b\in \theta_i(\nit{body}(\sigma))\}}(\nit{level}(b) + 1)$. We refer to the chase with atoms up to level $k$ as $\nit{chase}^k(\mc{P})$, while $\nit{chase}^{[k]}(\mc{P})$ is the instance constructed after $k \ge 0$ chase steps.

Note that, the chase steps are applied in a {\em level saturating} fashion, meaning that if there are more than one applicable rules, the one that has body atoms with smallest maximum level is applied. Also importantly, each pair of applicable rule and homomorphism is only applied once during the chase procedure.

The chase procedure stops if there is no applicable rule and assignment. The chase result, ${\it chase}(\mc{P})$ or ${\it chase}(D,\mc{P}^r)$ called the chase, is the result of the last chase step. If the chase procedure does not terminate, ${\it chase}(\mc{P})=\bigcup_{i=0}^\infty(I_i)$, in which, $I_0=D$, and, $I_i$ is the result of the i-th chase step for $i > 0$.


\section{Stickiness Property and its Generalization}\label{sec:app-stickiness}

In this section, we first formalize the {\em sch-property} introduced in~\cite{cali12} and we give an extension of it, {\em generalized stickiness property of the chase} ({\em gsch-property}). Both the {\em sch-property} and the {\em gsch-property} are defined based on the notions of the {\em chase relation} and the {\em chase derivation relation} that we explain here.

\begin{definition}\label{df:crel}\em Let $I_i\xrightarrow{\sigma_i,\theta_i}I_i\cup\{A_i\}$ be the $i$-th chase step of a program $\mc{P}$ that applies the rule $\sigma_i$ with $\theta_i$ as the assignment that makes the body of $\sigma_i$ true in $I_i$ and generates a new atom $A_i$. We define $\nit{rchase}(\mc{P})=\bigcup_{i=1}^{M}(\sigma_i[\theta_i] \times A_i)$ as the {\em chase relation} of $\mc{P}$, where $M$ is the minimum number of steps to make the chase stop (but $M=\infty$ if the latter does not stop). The {\em chase derivation relation} of $\mc{P}$, denoted by  $\nit{dchase}(\mc{P})$, is the transitive closure of $\nit{rchase}(\mc{P})$. \boxtheorem\end{definition}

Intuitively, $\nit{dchase}(\mc{P})$ contains every derivation of atoms in $\nit{chase}(\mc{P})$. In Example~\ref{ex:chase}, $\nit{dchase}(\mc{P})$ includes $(r(a,b),r(b,\zeta_1)),(r(a,b),s(a,b,\zeta_1))$ and $(r(a,b),r(\zeta_1,$ \ $\zeta_2))$.

\begin{definition} \label{df:sch}\em A program $\mc{P}$ has the {\em stickiness property} of the chase~\cite{cali12}, the {\em sch-property}, if and only if for every chase step $I_i\xrightarrow{\sigma_i,\theta_i}I_i\cup\{A_i\}$, the following holds: If a variable $X$ appears more than once in ${\it body}(\sigma_i)$, $\theta_i(X)$ occurs in $A_i$ and every atom $B$ for which, $(A_i, B) \in \nit{dchase}(\mc{P})$. \SCh \ is the class of programs with the {\em sch-property}. \boxtheorem \end{definition}

The concept of the {\em gsch-property} is specified by relaxing the condition for the {\em sch-property}: it applies only to values for repeated variables in the body of $\sigma_i$ that do not appear in so-called {\em finite positions} defined next.

\begin{definition} \label{df:finite} \em Given a program $\mc{P}$ with schema $\mc{R}$, the set of finite positions of $\mc{P}$, referred to as $\fp{\mc{P}}$, is the set of positions where finitely many values appear in $\nit{chase}(\mc{P})$. Every position that is not finite is infinite.\boxtheorem
\end{definition}

\begin{definition} \label{df:gsch}\em A program $\mc{P}$ has the {\em generalized-stickiness property of the chase} ({\em gsch-property}) if and only if for every chase step, $I_i\xrightarrow{\sigma_i,\theta_i}I_i\cup\{A_i\}$, the following holds: If a variable $X$ appears more than once in ${\it body}(\sigma_i)$ and {\em not} in $\fp{\mc{P}}$, $\theta_i(X)$ occurs in $A_i$ and every atom $B$ for which, $(A_i, B) \in \nit{dchase}(\mc{P})$. \GSCh \ is the class of programs with the {\em gsch-property}. \boxtheorem\end{definition}

\section{\ms} \label{sec:rewriting}

The \ms \ rewriting technique takes a \dplus \ program $\mc{P}$ and a CQ $\mc{Q}$ of schema $\mc{R}$ and returns a program $\mc{P}_m$ and a CQ $\mc{Q}_m$ of schema $\mc{R}_m$ such that $\nit{ans}_\mc{Q}(\mc{Q},\mc{P})=\nit{ans}_{\mc{Q}_m}(\mc{Q}_m,\mc{P}_m)$. Here we describe \ms \ in more details using the same program in Example~\ref{ex:mg}.

The rewriting uses the notion of {\em sideways information passing strategy} (\SIPS). A \SIPS \ of a rule specifies a propagation strategy in a top-down evaluation approach for the rule. Intuitively, a \SIPS \ of a rule is a strict partial order over the atoms of the rule which shows how the bindings are originated from the head and propagated through the body. 

\begin{definition} \em Let $p$ be a predicate of arity $k$. An {\em adornment} for $p$ is a string $\alpha=\alpha_1...\alpha_k$ defined over the alphabet $\{b,f\}$. The $i$-th argument of $p$ is considered {\em bound} if $\alpha_i=b$, or {\em free} if $\alpha_i=f$, $(1\le i\le k)$. The predicate $p^\alpha$ is an {\em adorned predicate} of $p$. Consider a \dplus \ rule $\sigma$ with a head predicate $p$ and an adornment $\alpha$ of $p$. Let $\nit{atoms}(\sigma)$ be the set of atoms in the body and the head of $\sigma$. A \SIPS \ of $\sigma$ and $\alpha$ is a pair $\langle<^{\sigma,\alpha},f^{\sigma,\alpha}\rangle$ in which $<^{\sigma,\alpha}$ is a strict partial order over $\nit{atoms}(\sigma)$ and $f^{\sigma,\alpha}$ is a function assigning to each atom $A \in \nit{atoms}(\sigma)$ an adornment such that $<^{\sigma,\alpha}$ and $f^{\sigma,\alpha}$ have the following properties:

\begin{enumerate}
\item For every atom $A \in \nit{body}(\sigma)$, $\nit{head}(\sigma)<^{\sigma,\alpha} A$.
\item $f^{\sigma,\alpha}(\nit{head}(\sigma)) = \alpha$.
\item If a variable $X$ in $A$ is bounded according to $f^{\sigma,\alpha}(A)$, $X$ either appears in $\nit{head}(\sigma)$ and it is bounded according to $f^{\sigma,\alpha}(\nit{head}(\sigma))$ or it occurs in an body atom $B \in \nit{body}(\sigma)$ such that $B <^{\sigma,\alpha} A$. Intuitively, this property says if a variable is bounded in an atom it is either bound in the head atom or it is already evaluated in a body atom.
\end{enumerate}
\noindent A \SIPS \ $\langle<_1^{\sigma,\alpha},f_1^{\sigma,\alpha}\rangle$ is included in a \SIPS \ $\langle<_2^{\sigma,\alpha},f_2^{\sigma,\alpha}\rangle$ iff for every atom $A \in \nit{\sigma}$ and variable $X \in A$, if a $A$ is bounded according to $f_1^{\sigma,\alpha}(A)$ it is bounded according to $f_2^{\sigma,\alpha}(A)$. A \SIPS \ is {\em partial} if it is included in another \SIPS \ and otherwise it is {\em full}.
\boxtheorem\end{definition}

Intuitively, a \SIPS \ is partial if it does not always propagate all available information. In Section~\ref{sec:magic} and specifically in Theorem~\ref{th:closed}, we consider full \SIPS. We discuss about \ms\ with a partial \SIPS \ in Section~\ref{sec:disc}.

\begin{example} (ex. \ref{ex:mg} cont.) \label{ex:app-ms} For the rule $\sigma: r(X,Y),r(Y,Z) ~\rightarrow~ p(X,Z)$ and the adornment $\alpha=\nit{bf}$, a possible \SIPS \ is $<^{\sigma,\nit{bf}}=\{(p(X,Z),r(X,Y)),(p(X,Z),r(Y,Z)),$ \ $(r(X,Y),r(Y,Z))\}$ and $f^{\sigma,\nit{bf}}=\{(p(X,Z),\nit{bf}),(r(X,Y),\nit{bf}),(r(Y,Z),\nit{bf})\}$.

This \SIPS \ is complete. A possible partial \SIPS \ for $\sigma$ and $\alpha$ is: $<_\nit{par}^{\sigma,\nit{bf}}=<_\nit{par}^{\sigma,\nit{bf}}$ and $f_\nit{par}^{\sigma,\nit{bf}}=\{(p(X,Z),\nit{bf}),(r(X,Y),\nit{bf}),(r(Y,Z),\nit{ff})\}$ in which for $f_\nit{par}^{\sigma,\nit{bf}}(r(Y,Z))$ both positions are free unlike $f^{\sigma,\nit{bf}}(r(Y,Z))$ with the first position bounded.
\boxtheorem\end{example}

\ms \ starts from the body atoms of $\mc{Q}$ and generates their adorned atoms by annotating their predicates with strings of $b$'s and $f$'s in the positions that contain constants and variables resp. We make a set of predicates $P$ with two types of adorned predicates: marked and unmarked. We add the new predicates of $Q$ into $P$ as unmarked predicates. Then we iteratively pick an unmarked predicate $p^\alpha$ from $P$ and generate its adorned rules and mark it as processed. For $p^\alpha$, we find every rule $\sigma$ with the head predicate $p$ and we generate an adorned rule $\sigma'$ as follows. We choose a \SIPS \ of $\sigma$ and $\alpha$ and we replace every body atom in $\sigma$ with its adorned atom and the head of $\sigma$ with $p^\alpha$. The adornment of the body atoms is obtained from the \SIPS \ and its function $f^{\sigma,\nit{bf}}$. If the generated adorned predicates from the body of $\sigma$ are not in $P$ we add them into $P$ as unmarked predicates. We add the adorned rule $\sigma'$ into $\mc{P}^r$ and after repeating this for every rule $\sigma$ we mark $p$.

\begin{example} (ex. \ref{ex:mg} cont.) For the CQ $p(a,Y)~\rightarrow~\nit{ans}_\mc{Q}$, its adorned rule is $p^{bf}(a,Y)$ \ $~\rightarrow~\nit{ans}_\mc{Q}$ which adds $p^{bf}$ to $P$. Adorning $r(X,Y),r(Y,Z) ~\rightarrow~ p(X,Z)$ with the head predicate $p^\nit{bf}$ results into an adorned rule $r^{bf}(X,Y),r^{bf}(Y,Z) ~\rightarrow~ $ \ $p^{bf}(X,Z)$ that  we add into $\mc{P}^r_m$. We add $r^{bf}$ to $P$ and mark $p^\nit{bf}$ as processed. Next, $r^{bf}$ results into the adorned rule $u(Y),r^{fb}(X,Y) ~\rightarrow~ \exists Z\;r^{bf}(Y,Z)$ and adds $r^{fb}$ into $P$ and marks $r^{bf}$ as processed. But, there is no adorned rule for $r^{fb}$ since $u(Y),r(X,Y) ~\rightarrow~ \exists Z\;r(Y,Z)$ can not be bounded in the position of the variable $Z$. The result set of adorned rule is:

\vspace{-3mm}
\[ \arraycolsep=0pt
\begin{array}{rl c rl}
\hspace{-8mm}r^{bf}(X,Y),r^{bf}(Y,Z) ~\rightarrow&~ p^{bf}(X,Z). &\hspace{8mm}& u(Y),r^{fb}(X,Y) ~\rightarrow&~ \exists Z\;r^{bf}(Y,Z).
\end{array}
\]

\vspace{-7mm}

\boxtheorem\end{example}

Now, for every adorned rule $\sigma'$ in $\mc{P}^r$ with the adorned head predicate $p^\alpha$, we add to the body of $\sigma'$ a magic atom with predicate $m\_p^\alpha$. The arity of $m\_p^\alpha$ is the number of occurrences of $b$ in the adornment $\alpha$, and its variables correspond to the bound variables of head atom of $p^\alpha$.

The magic predicates are defined by the magic rules constructed as follows. For every occurrence of an adorned predicate $p^\alpha$ in an adorned rule $\sigma'$, we construct a magic rule $\sigma''$ that defines $\nit{mg}\_p^\alpha$ (a magic predicate might have more than one definition).  We assume that the atoms in $\sigma'$ are ordered according to the partial order in the \SIPS \ of $\sigma$ and $\alpha$. If the occurrence of $p^\alpha$ is in atom $A$ and there are  $A_1,...,A_n$ on the left hand side of $A$ in $\sigma'$, the body of $\sigma''$ contains $A_1,...,A_n$ and the magic atom of $A$ in the head. We also create a seed for the magic predicates, in the form of a fact, obtained from the query.

\begin{example} (ex. \ref{ex:mg} cont.) Adding the magic atom $\nit{mg}\_p^{bf}$ to the adorned rule $r^{bf}(X,Y),$ \ $r^{bf}(Y,Z) ~\rightarrow~ p^{bf}(X,Z)$ we obtain $\nit{mg}\_p^{bf}(X), r^{bf}(X,Y),r^{bf}(Y,Z) ~\rightarrow~ $ \ $p^{bf}(X,Z)$. Similarly the adorned rule $u(Y),r^{fb}(X,Y) ~\rightarrow~ \exists Z\;r^{bf}(Y,Z)$ becomes $\nit{mg}\_r^{bf}(Y),u(Y),r^{fb}(X,Y) \rightarrow \exists Z\;r^{bf}(Y,Z)$. The following are the magic rules that define $\nit{mg}\_p^{bf}$ and $\nit{mg}\_r^{bf}$ (the seed atom for the magic predicates is, $\nit{mg}\_p^{bf}(a)$):

\vspace{-4mm}
\[ \arraycolsep=0pt
\begin{array}{rl c rl}
\hspace{-8mm}\nit{mg}\_p^{bf}(X) ~\rightarrow&~ \nit{mg}\_r^{bf}(X). &\hspace{8mm}& \nit{mg}\_r^{bf}(X),r^{bf}(X,Y) ~\rightarrow&~ \nit{mg}\_r^{bf}(Y).
\end{array}
\]

\vspace*{-7mm}
\boxtheorem\end{example}

$\mc{P}$ is a \dplus \ program that might have intentional predicates with extensional data in $D$. Therefore, we add rules to load the data from $D$  when such a predicate gets adorned. In the Example~\ref{ex:mg}, $r$ is an intentional predicates with the extensional data $r(a,b)$ and so we add the following to load this data into the adorned predicates $\nit{mg}\_r^\nit{bf}$ and $\nit{mg}\_r^\nit{fb}$:

\[ \arraycolsep=0pt
\begin{array}{rl c rl}
\hspace{-8mm}\nit{mg}\_r^{bf}(X),r(X) ~\rightarrow&~ r^{bf}(X). &\hspace{8mm}& \nit{mg}\_r^{fb}(X),r(X) ~\rightarrow&~ r^{fb}(X).
\end{array}
\]

\vspace{-6mm}

\section{Examples} \label{sec:examples}

\begin{example} \label{ex:wsch} Consider a program $\mc{P}$ with $D=\{r(a,b)\}$ and the following rules:
\begin{align}
r(X,Y) &\rightarrow \exists Z\;r(Y,Z).\\
c(X),r(X,Y),r(Y,Z) &\rightarrow u(X,Z).\label{eq:s5}
\end{align}
$\mc{P}$ is not \WS \ because $Y$ in (\ref{eq:s5}) is marked and does not appear in $\Pi_F(\mc{P}^r)$. The program is \WSCh \ because (\ref{eq:s5}) is never applied during the chase of $\mc{P}$. \boxtheorem
\end{example}


\begin{example} \label{ex:notclosed} Consider a program $\mc{P}$ with a database $D=\{r(a,b),v(b)\}$, a BCQ $\mc{Q}:r(Y,a)~\rightarrow~\nit{ans}_\mc{Q}$ and the following set of rules $\mc{P}^r$:
\begin{align}
r(X,Y) ~\rightarrow~& \exists Z\;r(Y,Z).\label{eq:mc1}\\
r(X,Y) ~\rightarrow~& \exists Z\;r(Z,X).\label{eq:mc2}\\
r(X,Y),r(Y,Z),v(Y) ~\rightarrow~& r(Y,X).\label{eq:mc3}
\end{align}

The program is \WS \ since the only repeated marked variable is $Y$ in (\ref{eq:mc3}) and it appears in $v[1] \in \Pi_F(\mc{P}^r)$. The marked variables are specified by a hat sign. The result of the magic-sets rewriting $\mc{P}^m$ is the following, with the adorned rules:
\begin{align}
r^\nit{fb}(Y,a)~\rightarrow~\nit{ans}_\mc{Q}.\\
\nit{mg}\_r(Y),r^\nit{fb}(X,Y) &\rightarrow \exists Z\;r^\nit{bf}(Y,Z).\label{eq:m1}\\
\nit{mg}\_r(X),r^\nit{bf}(X,Y) &\rightarrow \exists Z\;r^\nit{fb}(Z,X).\label{eq:m2}\\
\nit{mg}\_r(X),r^\nit{bf}(X,Y),r^\nit{bf}(Y,Z),v(Y) &\rightarrow r^\nit{fb}(Y,X).\label{eq:m3}\\
\nit{mg}\_r(Y),r^\nit{fb}(X,Y),r^\nit{bf}(Y,Z),v(Y) &\rightarrow r^\nit{bf}(Y,X).\label{eq:m4}
\end{align}
and the magic rules:
\begin{align}
&\nit{mg}\_r(a).\label{eq:m5}\\
\nit{mg}\_r(X),r^\nit{bf}(X,Y) ~\rightarrow~& \nit{mg}\_r(Y).\label{eq:m6}\\
\nit{mg}\_r(Y),r^\nit{fb}(X,Y) ~\rightarrow~& \nit{mg}\_r(X).\label{eq:m7}
\end{align}

Here, every body variable is marked. Note that according to the description of \ms \ in Appendix~\ref{sec:magic}, the magic predicates $\nit{mg}\_r^\nit{fb}$ and $\nit{mg}\_r^\nit{bf}$ are equivalent and so we replace them with a single predicates, $\nit{mg}\_r$.

$\mc{P}_m$ is {\em not} \WS, since $r^\nit{fb}[1], r^\nit{fb}[2], r^\nit{bf}[1], r^\nit{bf}[2],$ and $\nit{mg}\_r[1]$ are not in $\Pi_F(\mc{P}^r_m)$ so; (\ref{eq:m1}), (\ref{eq:m2}), (\ref{eq:m6}) break the syntactic property of \WS. Following the chase of $\mc{P}_m$, the program is not in \SCh$(\mc{S}^\nit{rank})$ either. That is because in (\ref{eq:m6}) $a$ replaces $X$ that appears only in infinite rank positions $\nit{mg}\_r[1]$ and $r^\nit{bf}[1]$.

$\mc{P}_m$ is \JWS. That is because, $r^\nit{fb}[2], r^\nit{bf}[1]$ are in $\Pi^\exists_F(\mc{P}^r_m)$ and every repeated marked variable appears at least once in one of these two positions which means $\mc{P}_m$ is \JWS. Note that both $r^\nit{fb}[2], r^\nit{bf}[1]$ are bounded positions and are finite-existential which confirms the first claim in the proof of Theorem~\ref{th:closed}.\boxtheorem\end{example}


\begin{example} \ignore{(ex. \ref{ex:notclosed})} \label{ex:notclosed-jws} In Example~\ref{ex:notclosed}, we applied full \SIPS s, that passe full information about the bounded variables during the evaluation of a rule. Here, we consider partial \SIPS s that generate the following program:

\vspace{-3mm}
\begin{align}
r^\nit{fb}(Y,a)~\rightarrow~\nit{ans}_\mc{Q}.\\
\nit{mg}\_r^\nit{ff},r^\nit{ff}(X,Y) &\rightarrow \exists Z\;r^\nit{ff}(Y,Z).\label{eq:p1}\\
\nit{mg}\_r(Y),r^\nit{ff}(X,Y) &\rightarrow \exists Z\;r^\nit{bf}(Y,Z).\label{eq:p2}\\
\nit{mg}\_r(X),r^\nit{ff}(X,Y) &\rightarrow \exists Z\;r^\nit{fb}(Z,X).\label{eq:p3}\\
\nit{mg}\_r(X),r^\nit{bf}(X,Y),r^\nit{bf}(Y,Z),v(Y) &\rightarrow r^\nit{fb}(Y,X).
\end{align}
and the magic rules:

\vspace{-5mm}
\begin{align}
&\nit{mg}\_r(a).\\
\nit{mg}\_r(X),r^\nit{bf}(X,Y) ~\rightarrow~& \nit{mg}\_r(Y).\\
\nit{mg}\_r(Y),r^\nit{fb}(X,Y) ~\rightarrow~& \nit{mg}\_r(X).
\end{align}

Specially, in (\ref{eq:p1})-(\ref{eq:p3}) the information about the bounded variables from the head atom is not used in the body. In (\ref{eq:p2}) and (\ref{eq:p3}), $\nit{mg}\_r[1]$ and $r^\nit{ff}$ are infinite positions and if we follow the chase, there are values that replace the join variables $Y$ and $X$ in these rules and the values do not propagate all the way to the head atoms in the next steps. Therefore, the result program is not \GSCh. This shows that using partial \SIPS, \GSCh \ or any of its semantic subclasses of \SCh$(\mc{S})$ are not closed under \ms.\boxtheorem\end{example}

\section{Discussion}\label{sec:disc}

\subsection{Connection with Partial Grounding Approach}

A new hybrid approach for QA over \WS \ programs is proposed in~\cite{milani16amw,milani16er}. In this approach, a given \WS \ program is rewritten by partially grounding some variables and transforming the program into a sticky program w.r.t the extensional data. An input CQ then is combined with the result sticky program to obtain a UCQ to be answered directly on the extensional database.

This hybrid approach that combines bottom-up grounding and backward rewriting is a new promising technique for QA. However, it strongly relies on the syntactic properties of the \WS \ and sticky programs. The QA algorithm in this paper applies for a range of semantic programs with certain property of their chase instance rather than specific syntactic properties.

\subsection{\ms \ with Partial Sideways Information Passing Strategies}\label{sec:notclosed}

In Section~\ref{sec:magic} and Theorem~\ref{th:closed}, we assumed that the \ms \ always uses full SIPS for generating the adorned rules and takes advantage of full information about the bound and free variables of the head atom and the already evaluated atoms in the body. This is specifically necessary to prove the claim that in the result of \ms \ every bounded position is finite-existential and so this is required to prove that the class of \JWS \ program is closed under \ms.

Example~\ref{ex:notclosed} in Appendix~\ref{sec:examples} shows a situation when using partial SIPSs in \ms \ and rewriting a \JWS \ program, the result is not \JWS. This is not a syntactic incident because the result is not even \GSCh. It means there is no semantic or syntactic subclass of \GSCh \ that uses a proper selection function and is closed under \ms.

However, this is not problematic for the integration of \QAA \ and \ms \ with partial SIPSs.  That is because the \QAA \ algorithm is still applicable for the result program with a modification in the applicability condition of in Definition~\ref{df:app}. Specifically, the adorned rules without their magic predicates for a set of rules that still preserve the stickiness. Therefore, we can ignore the condition (b) in Definition~\ref{df:app} for the magic rules and obtain a new QA algorithm that freely propagates the data of the magic predicates. With this modification the result algorithm still terminates since the magic predicates do not invent new values and it returns correct answers.

\subsection{Further Generalization of the Stickiness Property of the Chase}

In the {\em gsch-property}, we generalized the stickiness by relaxing the condition on the join variables when they appear at least once in a finite position. \GSCh \ (the class of program with the {\em gsch-property}) is an abstract class that can not be syntactically checked but it is an important class as it defines a decidability paradigm for the programs with different form of stickiness of the chase considering the \QAA \ algorithm that works for any sub class \SCh$(\mc{S})$ of \GSCh \ with computable $\mc{S}$.

Investigating \QAA \ and the proof of its correctness in Theorem~\ref{th:correctness}, we notice that we could even furhter relax the stickiness condition and define a more general class compared ti \GSCh. That is if a variable is replaced during the chase with a value that traversed a finite position at some point before reaching the current rule, we can relax the condition for the variable. This defines a more general class of programs compared to \GSCh \ that still can define decidable semantic classes with computable selection functions.

\end{document}